\documentclass[prb,preprintnumbers,amsmath,amssymb,twocolumn]{revtex4-2}

\usepackage{graphicx}
\usepackage{dcolumn}
\usepackage{bm}
\usepackage{appendix}
\usepackage{hyperref}
\usepackage[capitalise]{cleveref}
\usepackage{booktabs} 
\usepackage{multirow} 
\usepackage{subcaption} 
\usepackage{ragged2e}
\DeclareCaptionJustification{justified}{\justifying}
\usepackage{xspace}

\begin{document}
\title{Modeling neutron and X-ray scattering by liquids}

\author{Wouter Montfrooij$^1$, Ubaldo Bafile$^2$, and Eleonora Guarini$^3$} \affiliation{$^1$Department of Physics and Astronomy, and the Missouri Research Reactor, University of Missouri, Columbia, 65211 Missouri, United States,\\
$^2$Consiglio Nazionale delle Ricerche, Istituto di Fisica
Applicata "Nello Carrara", via Madonna del Piano 10, I-50019 Sesto Fiorentino, Italyy,\\
$^3$Dipartimento di Fisica e Astronomia, Universit\`{a} di Firenze, via G. Sansone 1, I-50019 Sesto Fiorentino, Italy.}
\begin{abstract}
{We review exact formalisms for describing the dynamics of liquids in terms of static parameters. We discuss how these formalisms are prone to suffer from imposing restrictions that appear to adhere to common sense but which are overly restrictive, resulting in a flawed description of the dynamics of liquids. We detail a fail-safe way for modeling the scattering data of liquids that is free from any unwarranted restriction and that models the scattering using the fewest possible number of free parameters. We also list some common habits in analyzing data and how these habits do not do justice to the accuracy of the results obtained in scattering experiments, and how these habits may stand in the way of rejecting some models used in describing the dynamics of liquids.}
\end{abstract}
\maketitle
\section{introduction}
When probing liquids by means of x-ray \cite{Burkel} or neutron scattering experiments \cite{Lovesey}, one gains information about the dynamical processes in the liquids that help relax density fluctuations back to equilibrium. This information is obtained in the form of the dynamic structure factor $S(q,\omega)$, a quantity that contains information about the response of a liquid when probed on length scales $\lambda = 2 \pi/q$, while transferring an amount of energy $E= \hbar \omega $ to the liquid. Here, $\hbar q$ stands for the momentum transferred to the liquid by the probing particle. In principle, the dynamic structure factor is proportional to the intensity of the scattered radiation; in practice, some data correction procedures need to be performed in order to get rid of unwanted scattering, such as scattering by the sample holder, or the unavoidable instances where multiple scattering events take place. In addition, spectrometers are set to accept a range of energy transfers and scattering angles, resulting in the smearing of the signal of interest. However, all the corrections that need to be carried out in order to obtain the desired dynamic structure factor are well documented, and in practice \cite{Heitmann} they do not represent an obstacle to obtaining $S(q,\omega)$ to a good approximation.\\

While the earliest neutron scattering experiments on liquids resulted in the determination of dynamic structure factors with large uncertainties, the phenomenal advances in neutron and x-ray scattering capabilities now allow for the determination of the dynamic structure factor with very high statistical accuracy, covering a dynamic range that even allows for the measuring of short-wavelength sound modes in metals. However, our methods of data analysis and model fitting are still rooted in the earlier days of scattering experiments where the main focus was on inferring propagation frequencies and relaxation rates of density fluctuations, rather than on distilling those quantities from experiment using the most restrictive models. In addition, our methods reflect that we had been used to only being able to probe a fairly limited dynamic range. As a historical accident, some of our methods and models are being applied to situations where they have been shown to be inappropriate. We aim to address this issue with this paper and present a fail-safe and model-independent method of analyzing scattering data.\\

\begin{figure}
\begin{center}              
\includegraphics*[viewport=50 60 520 560,width=80mm,clip]{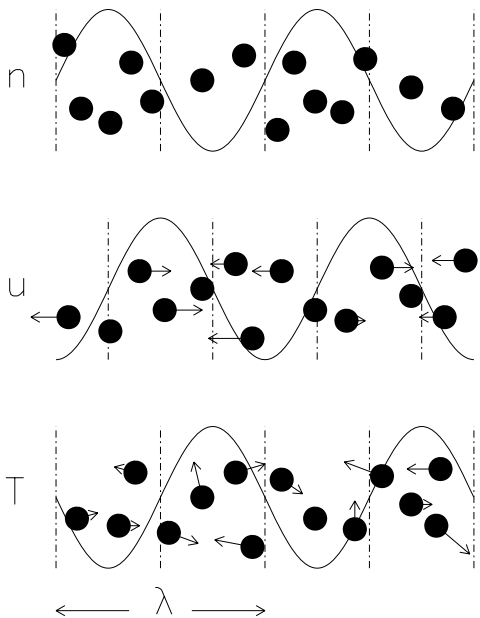}
\end{center}
\caption{A sketch representing microscopic fluctuations of wavelength $\lambda$ of the density  '$n$'
(top panel), velocity  '$u$' (middle panel) and temperature '$T$'  (bottom panel). In order for a density fluctuation to relax back to equilibrium, it must give rise to a longitudinal velocity fluctuation. This fluctuation in turn can relax through direct friction, or by giving rise to a fluctuation in the microscopic energy which can relax through binary collisions.} \label{decay}
\end{figure}

The dynamic structure factor of liquids is a measure of how spontaneous density fluctuations \cite{vanhove} relax back to equilibrium through coupling to other degrees of freedom in the liquid, such as momentum and energy fluctuations. In scattering experiments, we only probe the fate of density fluctuations directly \cite{Lovesey}; we have to infer all the couplings to the other fluctuations from these experiments. In order to do so, we employ models that reflect known restrictions, such as the fact that a density fluctuation necessarily must give rise to a longitudinal velocity (momentum) fluctuation in order to decay. This is shown in Fig. \ref{decay}.\\

In general, all fluctuations in liquids decay by coupling to other degrees of freedom (Fig. \ref{decay}). In order to assess the couplings mathematically, we use the Liouville operator $L$  \cite{balu,gnatzLJ} that describes the time evolution of a microscopic variable: $\frac{\partial}{\partial t} A(t) = iL A(t)$. We then look at the decay of the new microscopic variable $B(t)=\frac{\partial}{\partial t} A(t)$ generated by the action of the Liouville operator. We can repeat this process as many times as we like \cite{gnatzLJ}, until we reach a microscopic variable whose (spontaneous) fluctuations decay back to equilibrium so fast that we can replace the time dependence of the decay of this variable by its time integrated value. This is the principle behind the Mori-Zwanzig projection formalism \cite{zwanzig,mori}, the memory function formalism\cite{balu}, as well as the continued fraction expansion formalism\cite{balu}.\\

These formalisms allow for the separation of fluctuations in microscopic variables that decay (relatively) slowly, and those that decay very fast. Once this separation has been established, then we can describe the dynamics of liquids in terms of static quantities representing the couplings between microscopic variables (such as the coupling between the microscopic density and the microscopic longitudinal velocity\cite{gnatzLJ}) and the damping rates of fast decaying fluctuations, the latter representing the time-integrated values of a decaying fluctuation \cite{gnatzLJ}.\\

In principle, these formalisms are exact \cite{Forster,gnatzLJ,montfrooijEE,barocchi}, and apply to all liquids. As such, stringent restrictions are imposed on models aimed at describing the dynamics of liquids in terms of time-independent parameters. We explicitly write out the equivalent model expressions for these formalisms in this paper, however, the main point is that all these representations are equivalent, and as such they must abide by certain restrictions if they are to be used to describe the scattering by liquids. Representations that do not obey these restrictions fall in the category of phenomenological models. One of the goals of this paper is to convince the reader to restrict the usage of such phenomenological models, and instead to impose all the necessary restrictions where possible. Other goals are to ensure that the fitting procedures used in the analysis of scattering data are both as restrictive as possible, as well as that they do not impose model conditions that are not necessarily met in the dynamics of liquids.\\

While these formalisms are exact in principle, sometimes we encounter situations where our understanding of the Liouville operator is insufficient to state with certainty what microscopic variables couple to which ones. In particular, recently reported\cite{hosokawa,travmodes} propagating transverse modes in liquids cannot be incorporated directly into the projection formalism since there does not exist a direct coupling between longitudinal density fluctuations (pressure waves) and transverse propagating modes. In such cases, the data need to be modeled using a phenomenological model. In practice, existing models are extended by adding more terms, such as more Lorentzian lines to the projection formalism or more decay channels to the memory function.\\

Whether we use the exact formalisms or whether we use phenomenological models, all models need to adhere to certain restrictions. A perusal of the literature, as discussed in this paper, shows that it is not unusual for those restrictions to not be implemented in a rigorous enough manner, opening up the possibility of misinterpreting the information present in the scattering data. We present a fail-safe procedure for fitting scattering data that incorporates all restrictions and that can be applied in all cases, including the cases where our knowledge of the Liouville operator is insufficient. As such, Eq. \ref{failsafe}, to be discussed in the next section, is the main message of this paper.\\

The outline of the paper is as follows. We present all our arguments using monatomic liquids as an example; needless to say, our arguments apply to all liquids. In Section \ref{usethis} we describe the fail-safe method for analyzing scattering data. This method can be applied without any familiarity with the subsequesnt sections. Next, we review the basics of the projection formalism and its equivalent representations in Section \ref{formalism}. In Section \ref{phenomenon} we scrutinize some phenomenological models, and in Section \ref{analysis} we point out how we can routinely go wrong in doing our data analysis. In Section \ref{transverse} we discuss how to model excitations when the Liouville operator is not fully known. The latter is the case for transverse modes that appear to be visible in the dynamic structure factor as determined by scattering experiments and computer simulations.\\

As a last word in this introduction, there is absolutely no slight intended with this paper. Over the past decades some habits have slipped into our methods of data analysis, including the methods of the authors of this paper, that no longer do full justice
to the accuracy of the scattering data. The list of references to illustrate certain points is solely based on the familiarity of the authors with those papers, and on the results of a literature search.

\section{Model-free recipe to analyze scattering data}\label{usethis}

The quantity we access in scattering experiments \cite{vanhove} is the dynamic structure factor $S(q,\omega)$, which reflects the fate of density fluctuations that arise in a liquid. The quantity we are interested in is the dynamic susceptibility $\chi(q,\omega)$. The poles of the (complex) dynamic susceptibility give the excitations of the liquid. The two functions are related \cite{balu} through the fluctuation-dissipation theorem:

\begin{equation}
  \chi"(q,\omega)=\cfrac{1-e^{-\beta \hbar \omega}}{2\hbar}
S(q,\omega), \label{fluc1}
\end{equation}
where $ \chi"(q,\omega)$ (odd in $\omega$) is the imaginary part of the dynamic susceptibility. Here, $\beta$ is the inverse temperature $\beta = 1/k_BT$, with $k_B$ Boltzmann's constant while $\hbar$ denotes the reduced Planck constant $h/2\pi$. Thus, we do not have direct access to the full dynamic susceptibility: if we wish to obtain the real part of the susceptibility, we can do so indirectly by using the Kramers-Kronig relationships \cite{balu}; however, this procedure is not very accurate when dealing with scattering data. In addition, the frequency prefactor in Eq. \ref{fluc1} can skew the scattering results to higher frequencies, giving the impression that excitations propagate at higher speeds than they actually do, or even giving the impression that non-propagating modes propagate \cite{prbw}.\\

The central function in analyzing scattering data is the Fourier transform  $S_{sym}(q,\omega)$ of the relaxation function, defined by the following symmetrization of the dynamic structure factor:

\begin{equation}
S_{sym}(q,\omega)=\cfrac{1-e^{-\beta \hbar \omega}}{\beta
\hbar \omega} S(q,\omega) = \cfrac{2}{\beta \omega}\chi"(q,\omega).\label{sym2}
\end{equation}

It is this symmetrized function that is modeled in fits to the scattering data so that the dynamics of the liquid can be captured in terms of time-independent parameters such as the frequency moments of the dynamic susceptibility. The odd frequency moments of  $S_{sym}(q,\omega)$ equal zero, while the even frequency moments are related to the odd frequency moments of the dynamic structure factor  $S(q,\omega)$ as\cite{wouterbook}

\begin{equation}
\int_{-\infty}^{\infty}\omega^{2m} S_{sym}(q,\omega)d\omega = \cfrac{2}{\hbar\beta}\int_{-\infty}^{\infty}\omega^{2m-1} S(q,\omega)d\omega.\label{moments}
\end{equation}

Thus, the zeroth frequency moment of  $S_{sym}(q,\omega)$ is given by the static susceptibility $\chi_q$, whereas the static structure factor $S(q)$ [that is, zeroth frequency moment of  $S(q,\omega)$] does not correspond to a moment of  $S_{sym}(q,\omega)$. The second frequency moment of $S_{sym}(q,\omega)$  is determined by the f-sum rule [or equivalently, by the first frequency moment \cite{balu} of  $S(q,\omega)$]. We return to these distinctions in section \ref{analysis}.\\

 Barocchi and Bafile \cite{barocchi}  have shown that any correlation function can always be expressed as the sum of Lorentzians. Thus, not mereley can the density-density correlation function  be expressed as a sum of Lorentzian contribution, but any correlation function between two microscopic variables can be expressed by such a summation. This can be done even in the cases where we (expect to) see new excitations such as propagating transverse modes. We show in the remainder of the paper that the following equation provides a failsafe method of fitting scattering data, even in the absence of any knowledge about the liquid. When we fit our data to an odd number of Lorentzian lines then the fitting expression is given by: 

\begin{equation}
S_{sym}(q,\omega)=\cfrac{1}{\pi}\sum_i  \cfrac{A_i+B_i\omega^2}{(\omega^2-\Omega_i^2)^2+z_i^2\omega^2}+ \cfrac{1}{\pi} \cfrac{C \Gamma}{\omega^2+\Gamma^2}
\label{failsafe}
\end{equation}

When we fit to an even number of Lorentzian lines then the last term is absent from the fitting function. 
In this equation, each term in the summation represents one pair of Lorentzian lines characterized by four parameters $A_i$, $B_i$, $\Omega_i$, and $z_i$. The parameters $C$. $\Gamma$, $A_i$, $B_i$, $\Omega_i$, and $z_i$ are all \underline{real-valued} fitting parameters. The sought-after frequencies of the modes are determined by the $\Omega_i$ and $z_i$: we have a pair of propagating modes if $\Omega_i > z_i/2$ (with damping rate of $z_i/2$ and propagation frequencies of $\sqrt{\Omega_i^2-z_i^2/4}$), and we have two non-propagating modes if   $\Omega_i < z_i/2$ (with two different damping rates $\Gamma_{1,i}$ and $\Gamma_{2,i}$ determined by $\Omega_i=\Gamma_{1,i}\Gamma_{2,i}$ and $z_i$ = $\Gamma_{1,i} + \Gamma_{2,i}$). Only after the fitting procedure will it become known if, for each pair, we are looking at propagating modes, or non-propagating ones. We will return to these observations in later sections. In the case of an odd number of Lorentzian lines the last term in Eq. \ref{failsafe} is a central (in energy) diffusive (non-propagating) contribution. \\

Before fitting any particular model to scattering data, we suggest to follow the procedure below first in order to determine the minimum number of terms that provides a good fit to the scattering data. This ensures that we do not try to analyze our data with too many modes (projection formalism), decay channels (memory function formalism), or coupled Lorentzian lines (continued fraction expansion) during the more detailed stages of our data analysis:\\
\begin{enumerate}
\item Start with a small number of Lorentzian lines (two or three) using Eq. \ref{failsafe}.
\item Multiply the model with the frequency dependent prefactor shown in Eq. \ref{sym2}.
\item Numerically convolute the model with the spectrometer resolution function when needed (that is, when the resolution width is comparable to the intrinsic widths of the Lorentzian lines).
\item Determine the Lorentzian parameters as well as the overall scaling factor from a least-squares fitting procedure of the convoluted model to the data that have been corrected for multiple scattering.
\item Increase the number of Lorentzian lines by one and repeat the steps above and keep on increasing this number until the quality of the fit no longer improves.
\end{enumerate}

Following the above recipe it will become clear how many Lorentzian lines are needed. This number determines the dimension of the matrix used in the effective eigenmode description (projection formalism), the number of decay channels in the memory function formalism, or the number of fractions in the continued fraction expansion.\\

The number of parameters to be fitted are two per Lorentzian line plus one for the overall scale factor. However,
the number of parameters to be fitted to can be reduced by imposing frequency sum rules, such as the zeroth and second frequency moment of  $S_{sym}(q,\omega)$ that yield the static susceptibility and the f-sum rule, respectively. In practice, only the f-sum rule tends to be implemented. The existence of the f-sum rule implies the following restriction:

\begin{equation}
C \Gamma+\sum_i  B_i =0.
\label{dothis}
\end{equation}

This equation allows for the fitting of the data with one fewer free parameter. For instance, when fitting to a damped harmonic oscillator function (two Lorentzian lines, C= 0), then the coefficient B must equal zero in order to be able to satisfy the (exact) f-sum rule. Conversely, when fitting to three Lorentzian lines, the factor B cannot equal zero as this would imply the absence of the central line. We can still try to fit our data to a three-Lorentzian line expression with B= 0 \cite{hosokawa,hosokawa2}, but when we do then we know from the outset that our model will be phenomenological as it does not satisfy the exact f-sum rule and it will be unlikely that the fit will produce good agreement with the experimental data.\\

Requiring the existence of higher sum-rules imposes restrictions on the set \{$A_i$\}. Note that it is far from straightforward to include the static structure factor $S(q)$ as a fitting restriction since the sum of Lorentzian lines for  $S_{sym}(q,\omega)$ does not yield an analytic expression for the zeroth frequency moment of $S(q,\omega)$. Moreover, should the Liouville operator be known, such as in the case for the projection formalism, then the number of free parameters is greatly reduced. We give an example on Section \ref{formalism}.A how the dynamics of a 5-Lorentzian model are capurued by only seven parameters (rather than 10).  Lastly, by doing an absolute normalization measurement of the experimental setup the overall scale factor can be removed from the list of adjustable parameters.

\section{Projection formalism and equivalent representations}\label{formalism}

As discussed in the previous section, the central function in the formalisms capturing the dyncamics of liquids is $S_{sym}(q,\omega)$. In this section we give some explicit examples for what this implies for analyzing scattering data employing the (equivalent) descriptions of the projection formalism, such as the effective eigenmode description, the continued-fraction description, the memory function formalism, as well as a sum of Lorentzians. We stress that all these descriptions are equivalent, and, in principle, exact when applied properly to $ S_{sym}(q,\omega)$. We added the word properly to the previous sentence; in Sections \ref{phenomenon} and \ref{analysis} we discuss cases where the models have been applied improperly. Note that  $S_{sym}(q,\omega)$ is already being used extensively in the literature, especially when analyzing scattering by liquid metals \cite{ScopignoReview}. We list the various equivalent representations in Table \ref{equivalent}.

\begin{table*}[t]
 {
  \caption{The equivalent expressions \cite{bafileoverview} for the projection formalism (sketches) in
  terms of continued fractions $H$ and the memory function $M(q,t)$. For brevity, the
  $q$-dependences of the quantities have not been displayed.}
 \label{equivalent}
 }
 {
  \begin{tabular*}{\textwidth}[t]{l @{\extracolsep{\fill}} l l} \toprule\toprule
   &   \\[-8pt]
   model sketch & 1/$H$ & $M(q,t)$\\[3pt]
      \midrule
   & \\[-10pt]
   \raisebox{-0.5\height}{\includegraphics*[viewport=00 -100 600 200,width=38mm,clip]{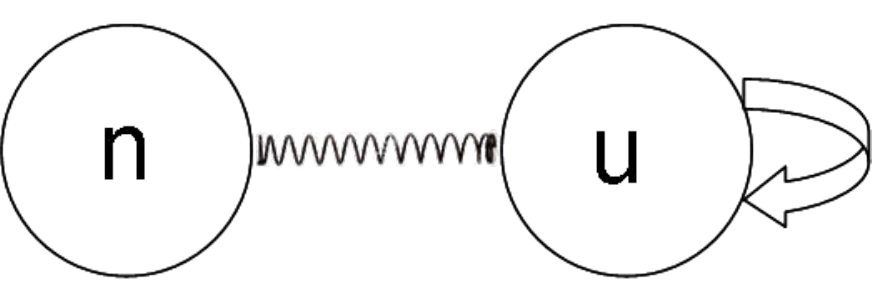}}
 & $\ i\omega+\cfrac{f_{un}^2}{i\omega+z_u} $  & $2z_u\delta(t)$\\[3pt]

 \raisebox{-0.6\height}{\includegraphics*[viewport=00 -100 600 200,width=38mm,clip]{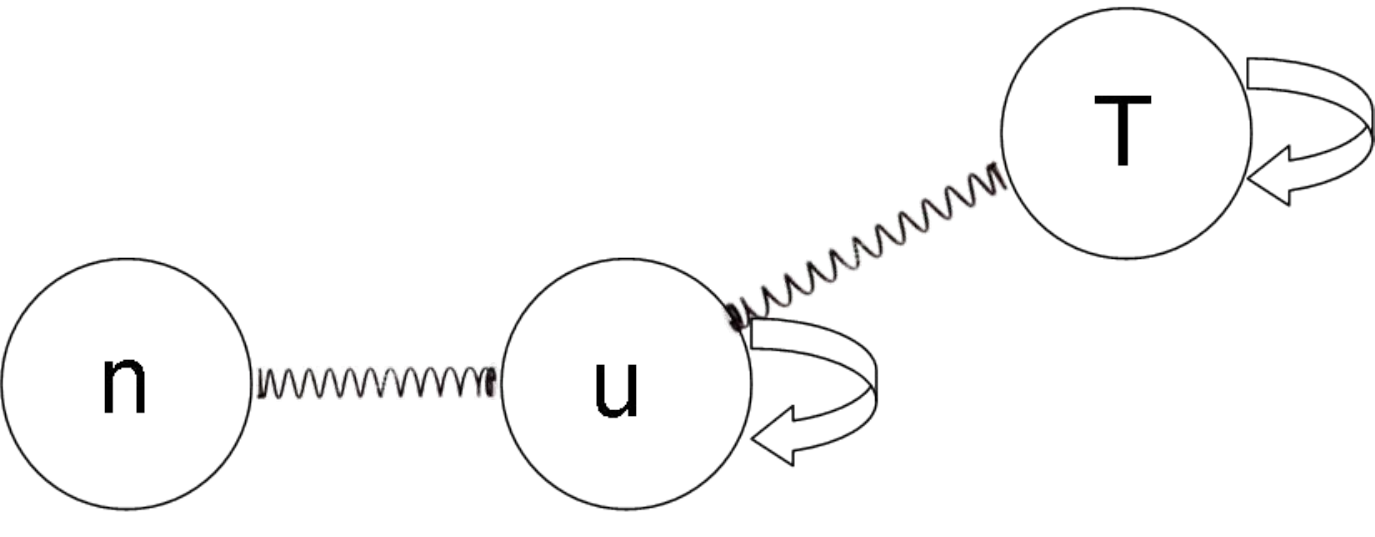}}
& $
i\omega+\cfrac{f_{un}^2}{i\omega+z_u+\cfrac{f_{uT}^2}{i\omega+z_T}} $ & $2z_u\delta(t)+f_{uT}^2e^{-z_Tt}$ \\[3pt]

\raisebox{-0.5\height}{\includegraphics*[viewport=00 -100 600 300,width=38mm,clip]{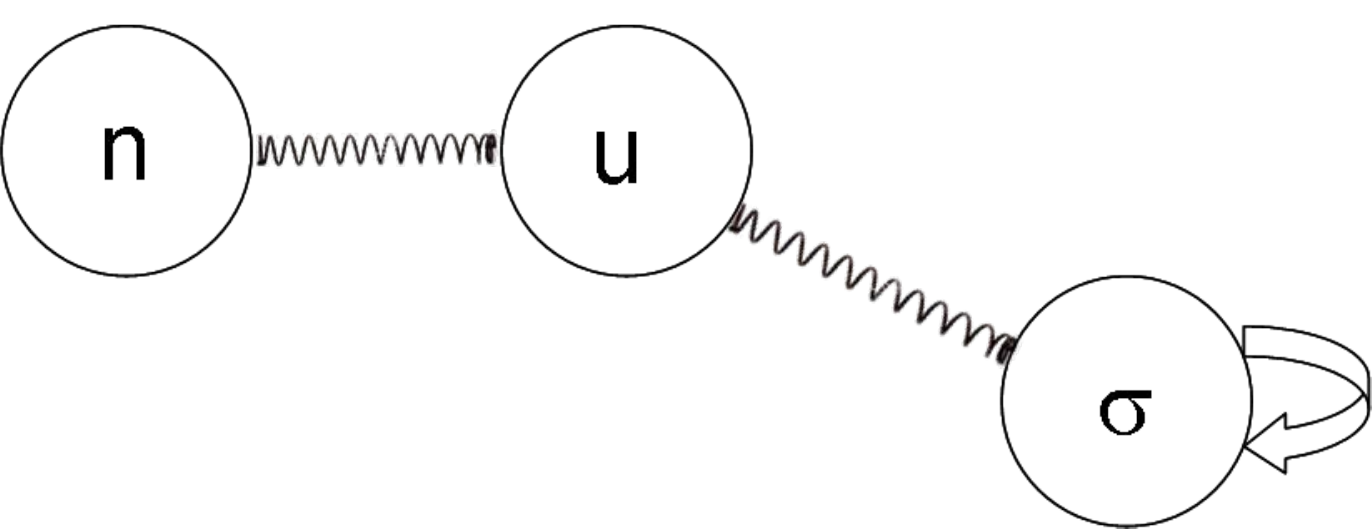}}
 & $ i\omega+\cfrac{f_{un}^2}{i\omega+\cfrac{f_{u\sigma}^2}{i\omega+z_{\sigma}}}$  & $f_{u\sigma}^2e^{-z_{\sigma}t}$\\[3pt]

\raisebox{-0.6\height}{\includegraphics*[viewport=00 -100 600 320,width=38mm,clip]{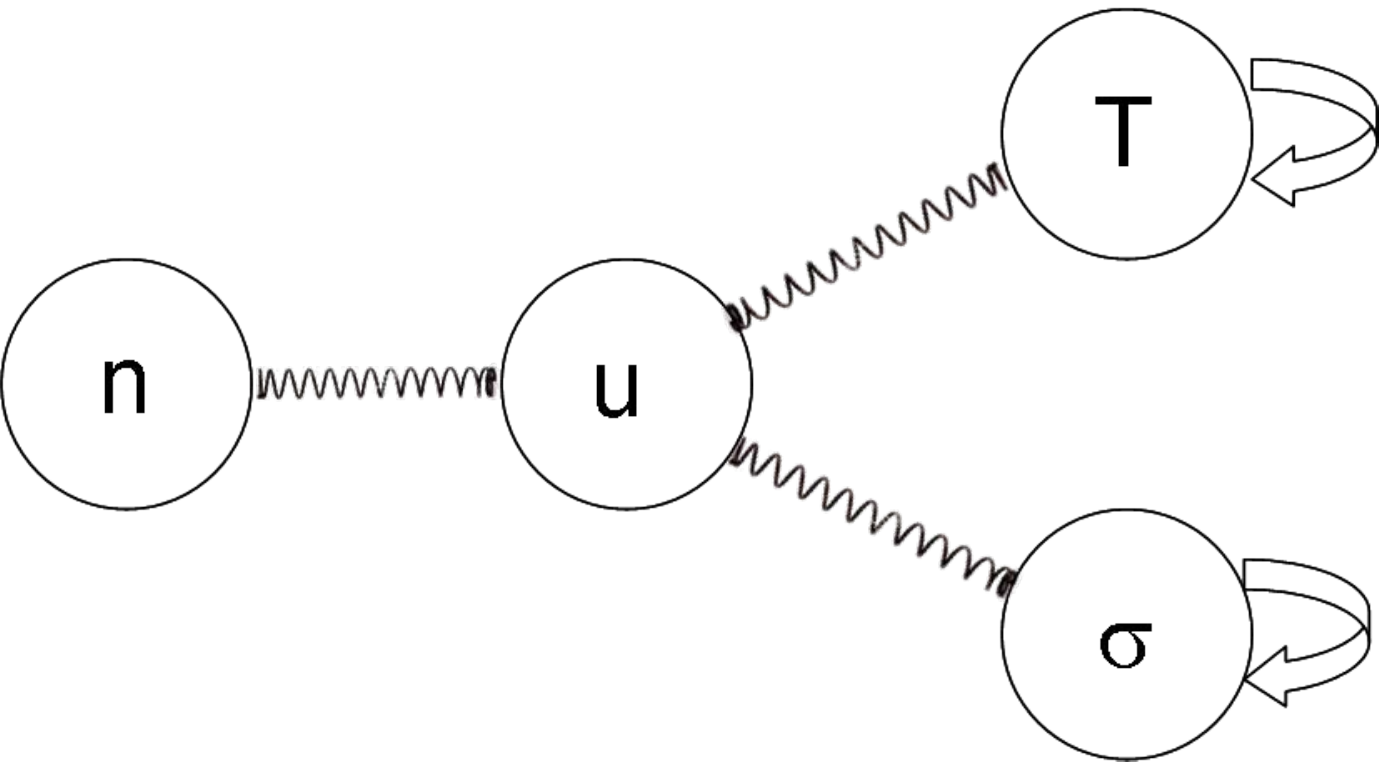}}
 & $\i\omega+\cfrac{f_{un}^2}{i\omega+\cfrac{f_{uT}^2}{i\omega+z_{T}}+\cfrac{f_{u\sigma}^2}{i\omega+z_{\sigma}}} $  & $f_{uT}^2e^{-z_Tt}+f_{u\sigma}^2e^{-z_{\sigma}t}$\\[3pt]

\raisebox{-0.6\height}{\includegraphics*[viewport=00 -100 600 330,width=38mm,clip]{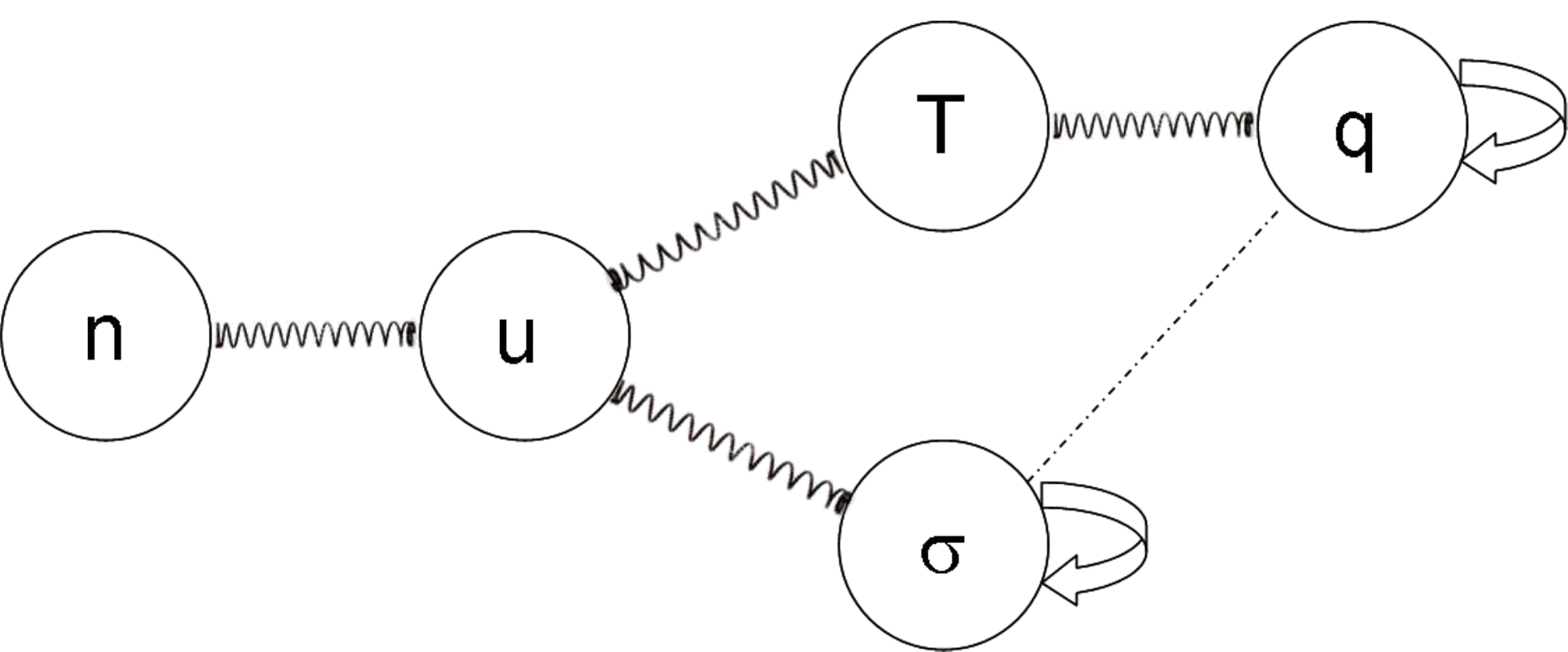}}
  & $ i\omega+\cfrac{f_{un}^2}{i\omega+\cfrac{f_{uT}^2}{i\omega+\cfrac{f_{Tq}^2}{i\omega+z_{q}}}+\cfrac{f_{u\sigma}^2}{i\omega+z_{\sigma}}} $ & $f_{uT}^2[se^{-rt}-re^{-st}]/(s-r)+f_{u\sigma}^2e^{-z_{\sigma}t}$;\\[-22pt]
($z_{q\sigma}= 0$)&& $s,r=z_q/2\pm[z_q^2/4-f_{Tq}^2]^{1/2}$ \\[3pt]
&&  \\[13pt]

\raisebox{-0.6\height}{\includegraphics*[viewport=00 280 600 550,width=38mm,clip]{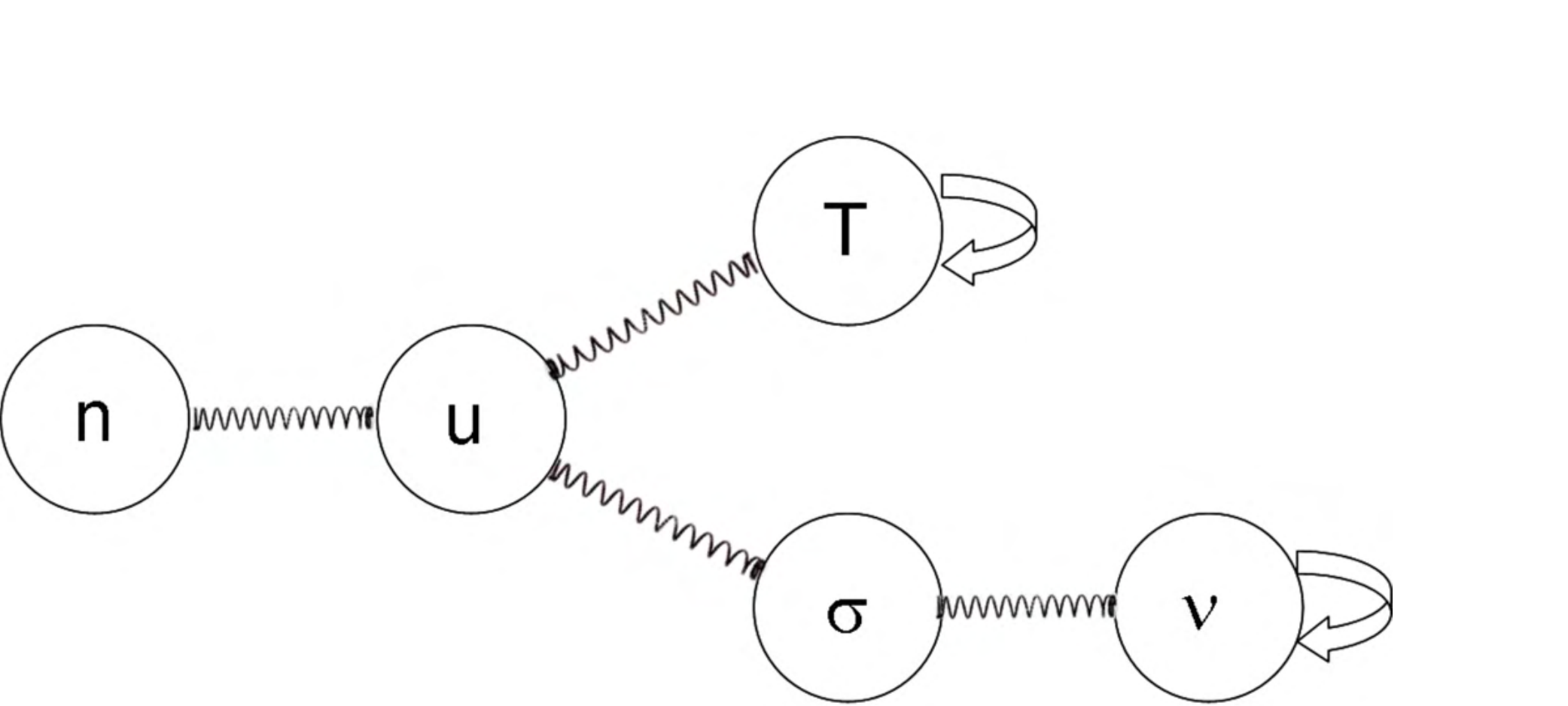}}
  & $\ i\omega+\cfrac{f_{un}^2}{i\omega+\cfrac{f_{u\sigma}^2}{i\omega+\cfrac{f_{\sigma\nu}^2}{i\omega+z_{\nu}}}+\cfrac{f_{uT}^2}{i\omega+z_{T}}} $ & $f_{u\sigma}^2[se^{-rt}-re^{-st}]/(s-r)+f_{uT}^2e^{-z_Tt}$;\\[-22pt]
&& $s,r=z_{\nu}/2\pm[z_{\nu}^2/4-f_{\sigma\nu}^2]^{1/2}$ \\[3pt]
&& \\[3pt]
\bottomrule\bottomrule
  \end{tabular*}
  }
\end{table*}

\subsection{Effective eigenmode formalism}

While the effective eigenmode formalism \cite{gnatzLJ} is not the oldest method speaking historically, it is the easiest to visualize in terms of diagrams. In order to assess how fluctuations in the microscopic density ultimately decay, we apply the Liouville operator $L$ to the microscopic variables, yielding a succession of  new microscopic variables. The Liouville operator yields the time-derivative of a microscopic variable $A(t)$ according to $iLA(t)= \partial A(t)/\partial t$. For instance, the Liouville operator applied to the microscopic density yields the longitudinal  microscopic velocity (momentum); the operator applied to the microscopic energy yields the microscopic energy flux. We stress that this holds for the case where the expression for the Liouville operator is known. Should the operator contain an unidentified term, such as a term that introduces rotations that could lead to a coupling to transverse modes, then we would lose track of the microscopic variables. We discuss this case in section \ref{transverse}. Until we reach that section, all that follows pertains to a Liouville operator that only depends on the position and velocities of all the particles in the liquid.\\

\begin{figure}
\begin{center}              
\includegraphics*[viewport=0 0 550 250,width=80mm,clip]{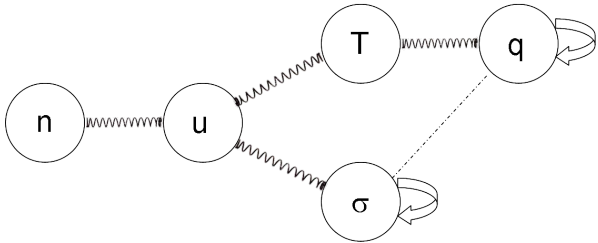}
\end{center}
\caption{Schematic diagram of how a density fluctuation '$n$' gives
rise to a velocity disturbance '$u$'. The velocity disturbance
relaxes back to equilibrium by causing a temperature disturbance
'$T$' and by setting up microscopic stress '$\sigma$'. The
microscopic stress gives rise to another microscopic disturbance
that relaxes so rapidly that its influence can be captured by a
decay rate (arrow). The temperature disturbance creates a
microscopic heat flux '$q$', which also gives rise to a microscopic
disturbance that decays very quickly. In here, the springs denote the
coupling constants $f_{ij}$ between the microscopic variables, and
the arrows looping back denote their decay rates. Corresponding to
this decay tree, there would be 4 coupling constants ($f_{un}$,
$f_{uT}$, $f_{u\sigma}$ and $f_{Tq}$), as well as two decay rates
($z_{\sigma}$ and $z_q$). The dotted line between '$q$' and
'$\sigma$' shows another allowed connection ($z_{q\sigma}$), even though in practice
 \cite{gnatzLJ} this connection has been found to
be very weak in the case of a Lennard-Jones fluid.} \label{allmodes}
\end{figure}

In applying the projection formalism, we determine how the density fluctuations decay by coupling to other variables such as the microscopic variables generated by the action of the Liouville operator. We show an example of such a procedure in Fig. \ref{allmodes}. In this figure we can see how a fluctuation decays by coupling to the various microscopic variables. Once we reach variables whose departures from equilibrium decay on time scales much faster than the time scales of interest to the experiment, then we can simply represent this decay rate by a number, rather than including the shape (in $\omega$) of the function governing the decay. The extent of the decay tree follows from model fits: should we not achieve a good fit to the data, then we simply extend the decay tree further. The advantage of this visualization (Fig. \ref{allmodes}) is that we can readily capture this decay tree in terms of a matrix $G(q)$ that includes all couplings between the microscopic variables, as well as the decay rates at the points where we cut off the tree. Explicitly, when translated to math, this paragraph cast in equations reads:
 \begin{equation}
 \begin{array}{ccl}
 S_{sym}(q,\omega)&=&\cfrac {1-e^{-\beta \hbar \omega}}{\beta\hbar \omega} S(q,\omega)\\
&=&
\cfrac{S_{sym}(q)}{\pi}Re \left[ \cfrac{1}{i\omega1
+G(q)}\right]_{11}. \end{array}
  \label{main}
  \end{equation}

In here, $S_{sym}(q)$ is given by the static susceptibility $\chi_q$ (not to be confused with the static structure factor $S(q)$, see section \ref{analysis}.D), and the label '11' denotes that we have to take the 11-element of the matrix $[i\omega1
+G(q)]^{-1}$. Furthermore, $'$Re$'$ stands for taking the real part of the expression while the matrix $G(q)$ yields all the couplings between the microscopic variables (such as the five shown in  Fig. \ref{allmodes}). For example, the matrix corresponding to the decay tree shown in Fig. \ref{allmodes}, and using the symbols 'n', 'u',' 'T', '$\sigma$', and 'q', respectively, is given by
   \begin{equation}
  G(q) = \left( \begin{array}{ccccc} 0 & if_{un}(q) & 0 &0 & 0\\
  if_{un}(q) & 0 &  if_{uT}(q) & if_{u\sigma}(q) & 0 \\
  0  &  if_{uT}(q) & 0 & 0 & if_{Tq}(q) \\
0 & if_{u\sigma}(q) & 0 & z_{\sigma}(q) & iz_{q\sigma}(q)\\
  0 & 0 & if_{Tq}(q) & iz_{q\sigma}(q) & z_q(q)\end{array}     \right).
  \label{m55}
  \end{equation}

Thus, it is straightforward to go from the depiction of any decay tree to a model that can be fitted to the scattering data with the forces (springs) being off-diagonal elements and the damping rates elements on the diagonal. The elements of the matrix $G(q)$ and the prefactor $S_{sym}(q)$ are then determined from the data through a fitting procedure employing Eq. \ref{main}.\\

Note that it is not necessary to invert the matrix $G(q)$ in order to fit the data. Since we only need the 11-element of the inverse matrix, we can fit to the ratio of two determinants (the minor of the 11-element divided by the full determinant). Thus, we would fit the data to
   \begin{equation}
S_{sym}(q,\omega)=
\cfrac{S_{sym}(q)}{\pi}Re 
\cfrac{\left| \begin{array}{cccc} 
  i\omega &  if_{uT}(q) & if_{u\sigma}(q) & .. \\
  if_{uT}(q) & i\omega & 0 & .. \\
 if_{u\sigma}(q) & 0 & i\omega+z_{\sigma}(q) & ..\\
.. &..&..&..\end{array}     \right|} 
{\left| \begin{array}{cccc} i\omega & if_{un}(q) & 0 & ..\\
  if_{un}(q) & i\omega &  if_{uT}(q) & .. \\
  0  &  if_{uT}(q) & i\omega & .. \\
.. & .. & .. & ..\end{array}     \right|} .
  \label{m56}
  \end{equation}
Computers are very good at calculating determinants, so Eq. \ref{m56} provides for a fast fitting routine, even when convoluting with the experimental resolution function.\\

In practice, one starts with a small matrix, such as a 3 x 3 matrix, and determines whether a matrix of this size can satisfactorily describe the scattering data. If not, one tries again with a larger matrix, ensuring of course that the matrix represents an allowed decay tree. One of the main advantages of using Eq. \ref{main} is that all the frequency sum rules that act as restrictions on the model parameters are automatically accounted for \cite{montfrooijEE} since the matrix was derived using the Liouville operator. Thus, we would be fitting our data using the fewest possible number of free parameters.

\subsection{Sum of Lorentzian lines}\label{lorentzian}
It is common practice to fit the observed dynamic structure factor to a sum of Lorentzian contributions. Some examples of this can be found in references [\onlinecite{gnatzAr,vanwellNe,vanwellAr,BoveHg,BoveK,BoveGa}]. For instance, in the hydrodynamic region where the longitudinal excitations are given by two propagating sound modes and one diffusive heat mode, we can represent $S(q,\omega)$ by the sum of three Lorentzian lines. Lorentzian lines that are centered around $\omega$= 0 are diffusive modes, lines that are centered around finite energies (frequencies) are propagating modes. The widths (in energy) of the Lorentzian lines are given by the decay rates of the particular modes, such as the decay rate of a propagating sound wave \cite{gnatzAr}. The matrix description of Eq. \ref{main} fully justifies this approach for any number of Lorentzian lines, not merely the three Lorentzian lines that we encounter in the hydrodynamic regime \cite{balu}. This can be seen as follows.\\

Since the values of the matrix elements of $G(q)$ fully determine the dynamic structure factor, they must also determine the eigenvectors of this matrix. The complex eigenvalues $\nu_i(q)$ of this matrix represent the poles of the dynamic susceptibility $\chi(q,\omega)$ \cite{wouterbook}. Rewriting Eq. \ref{main} in terms of these eigenvalues and eigenvectors $\psi_i$ we find
\begin{equation}
S_{sym}(q,\omega)=
\cfrac{S_{sym}(q)}{\pi}Re\sum_{i=1}^{m}\cfrac{[\psi^{(i)}_1(q)]^2}{i\omega+\nu_i(q)}.
\label{sum}
\end{equation}
In this expression, the square matrix has $m$ rows, and its $m$ eigenvectors are normalized according to \cite{gnatzLJ}
\begin{equation}
\sum_{k=1}^{m}\psi^{(i)}_k \psi^{(j)}_k= \delta_{ij}.
\end{equation}

Note that this orthonormalization does not involve taking the complex conjugates of the eigenvectors. Also note that Eq. \ref{sum} is still exact (provided L is known)  since it contains the same information as Eq. \ref{main}. Every real eigenvalue $\nu_i(q)$, representing a diffusive mode, has a real amplitude $[\psi^{(i)}_1(q)]^2$ in $S_{sym}(q,\omega)$, whereas every complex eigenvalue, representing propagating modes, corresponds to a complex amplitude. When we take the real part of a Lorentzian line with a complex amplitude, we obtain distorted (in energy) Lorentzian contributions \cite{bafileoverview} in $S_{sym}(q,\omega)$. More on this asymmetry in Section \ref{restrictive}.\\

One important thing to note is that the relative contributions of the Lorentzian lines in Eq. \ref{sum} to the dynamic structure factor are fully determined by the matrix $G(q)$. Therefore, the total number of free parameters  in the fitting procedure are still given by the number of parameters in the matrix $G(q)$ and the prefactor $S_{sym}(q)$; not all amplitudes can be treated as free parameters (as parameters independent of the positions and widths of the Lorentzian lines from our fit). It is for this reason that Eq. \ref{failsafe} tends to represent an initial fit with too many free parameters. As already mentioned, Barocchi and Bafile have shown \cite{barocchi} that any correlation function can always be expressed as the sum of Lorentzian lines, even in the case where the Liouville operator is not known exactly. Thus, Eq. \ref{sum} is valid even in the case when the amplitudes of the Lorentzian lines cannot be linked directly to a matrix of known structure forcing us to resort to fitting our data to Eq. \ref{failsafe}. We will use the latter method when discussing transverse modes that have been reported \cite{travmodes} in the dynamic structure factor of liquids.\\

\subsection{Continued-fraction expansion}
The continued fraction expansion is very closely related to Eq. \ref{main}. It simply involves taking the 11-element of the matrix $[i\omega1
+G(q)]^{-1}$ and writing it out explicitly. As such, this description is fully equivalent to those discussed before. While the details of the matrix description can be directly written down based on the visualization of decay trees (such as the one shown in Fig. \ref{allmodes}), we can in fact bypass the matrix description altogether and directly write down the equivalent continued-fraction expression following a recipe.\\

The conversion from the sketch of the decay tree to a mathematical expression is straightforward. Every microscopic variable (a circle with a letter in it in the sketch of a decay tree) yields a term '$i\omega + ...$' where the rest of the decay tree follows the '+' sign. Every spring (coupling constant) between two microscopic variables 'a' and 'b' gives a term '$f_{ab}^2/...$'. Again, the dots will be replaced with the remainder of the tree. Every arrow at the end of the tree, representing a decay rate, will give rise to a term '$z_c$ ', where the symbol 'c' stands for the particular microscopic variable. Lastly, when all the terms are collected, the expression is inverted to yield the continued-fraction expansion. For example, the visco-elastic decay tree shown in Fig. \ref{viscoelastic} is represented by the following continued-fraction expansion:
\begin{equation}
\left [ i\omega+\cfrac{f_{un}^2}{i\omega+\cfrac{f_{uT}^2}{i\omega+z_{T}}+\cfrac{f_{u\sigma}^2}{i\omega+z_{\sigma}}} \right ]^{-1}.
\label{above}
\end{equation}
In fitting to a model, the real part is then taken (by hand or by computer) of expressions similar to Eq. \ref{above}, the result is multiplied by the frequency factor implicit in Eq. \ref{main}, and compared to the experimental data points. In this way, all the static parameters that enter Eq. \ref{above} can be determined through a fitting procedure.

\subsection{Memory function formalism}

\begin{figure}
\begin{center}              
 \includegraphics*[viewport=0 0 400 240,width=55mm,clip]{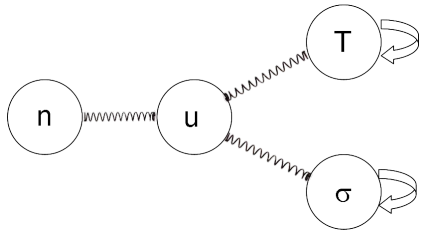}
\end{center}
\caption{The decay tree representative of the visco-elastic
model.} \label{viscoelastic}
\end{figure}

The memory function formalism has a long history \cite{balu} in the description of liquids, and it is the go-to method in the description of liquid metals \cite{ScopignoReview}. The memory function $M(q,\omega)$ describes the dynamic susceptibility by removing the part that is given by the exact f-sum rule \cite{Lovesey,balu}, that is, the part determined by $f_{un}(q)^2$ = $q^2/\beta m S_{sym}(q)$ in the notation used in this paper. By comparing this procedure to the continued-fraction expansion, we can readily identify the memory function as being the part '...'  in the denominator of the fraction '$f_{un}(q)^2/(i\omega+...$'. In particular, for the visco-elastic model of Eq. \ref{above} (shown in Fig. \ref{viscoelastic}), the memory function is given by:
\begin{equation}
M(q,\omega)=\cfrac{f_{uT}^2}{i\omega+z_{T}}+\cfrac{f_{u\sigma}^2}{i\omega+z_{\sigma}}.
\end{equation}

It is more common to represent this complex function in the time domain where it takes on real values \cite{ScopignoLia}. For the situation shown above, the memory function $M(q,t)$ in the time domain reads \cite{bafileoverview}:
\begin{equation}
M(q,t)=f_{uT}^2e^{-z_Tt}+ f_{u\sigma}^2e^{-z_{\sigma}t}.
\label{bla}
\end{equation}

Similar to the other representations of the projection formalism, the memory function formalism is exact, and the memory kernel, when identified through the equations
\begin{equation}
\begin{aligned}
 S_{sym}(q,\omega) &= \cfrac{S_{sym}(q)}{\pi}Re \left
[ \cfrac{1}{i\omega+G(q)} \right ]_{11}\\[10pt]
&= \cfrac{S_{sym}(q)}{\pi}Re \left
[ \cfrac{1}{i\omega+\cfrac{f_{un}(q)^2}{i\omega+M(q,\omega)}} \right
]_{11}\\[10pt]
&=\cfrac{1}{\pi}\cfrac{q^2/(\beta
m)M'(q,\omega)}{[\omega^2-f_{un}(q)^2-\omega M''(q,\omega)]^2+[\omega
M'(q,\omega)]^2},
\end{aligned}
\label{rewrite}
\end{equation}
contains all the relevant information regarding the poles of the dynamic susceptibility. In here, the real and the complex parts of the memory function $M(q,\omega)$ have been explicitly written out in order to make the comparison with the literature more straightforward. We list the memory functions for some commonly (and not so commonly) used models in Table I. Especially when the memory kernels are inspected in the time domain, it is clear why the memory function formalism is so widely used: the model parameters are directly identifiable as being part of a specific decay channel.\\

One drawback of the memory function approach is that it is not intuitively clear how important a certain decay channel is when it comes to the relaxation of density fluctuations. More concisely put, imagine a particle locked up in a cage formed by its neighbors. This particle will collide with its neighbors, until it eventually escapes its cage. This process is knows as cage diffusion. When modeling this process through the memory function formalism, we find that the decay channel whose strength in Eq. \ref{bla} is given by $f_{u\sigma}^2$ to be dominant. After all, this channel represents how the longitudinal momentum couples to the longitudinal momentum flux and decays. This process happens in essentially a single collision, since the particle will bounce back from the cage wall and the correlation with the value of its momentum before
 the collision will be largely lost. However, the density fluctuation itself, being the particle trapped in the cage, has not decayed at all since the particle is still inside the cage. In other words, this process would have very low amplitude in the dynamic structure factor. This is something we have to bear in mind when we perform our model fitting: processes that have very small amplitudes in $S(q,\omega)$ are difficult to extract, even when they can completely dominate the memory kernel.

\section{phenomenological models and restrictions on models}\label{phenomenon}
In the preceding section we have described the main formalisms used in the literature to analyze the dynamics of liquids. All formalisms are equivalent and, in principle, exact. Of course, when fitting to the data we must respect the restrictions built into the formalism at the risk of losing our exact description. For instance, when describing the scattering data by a sum of Lorentzians, the relative amplitudes of those Lorentzians can be restricted depending on the model that is being fitted to. In this section we list various inaccuracies that have slipped into our analysis of scattering data over the past half century. We point out where those methods fall short of being exact, and how they can be remedied. The overall aim is to ensure that we analyze our data using models that are as restrictive as possible. When we do the latter, we should be able to reject some models that otherwise may appear to give a good description of the data.\\

In general, there is no need to employ phenomenological models to fit scattering data of longitudinal excitations. In fact, the opposite holds true: when we fit our data to the most restrictive (exact) models for the decay of density fluctuations and we fail to obtain a satisfactory fit, then we can be sure that we have uncovered a new type of excitation, such as propagating transverse modes that show up in the longitudinal spectra. Also, we should bear in mind that fitting parameters of phenomenological models are just that: fitting parameters. These parameters are not guaranteed to represent the poles of the dynamic susceptibility. Unfortunately, some of the models that we think of as being exact (or have the potential of being exact) and that are believed to satisfy known restrictions, still turn out to be phenomenological models. The usage of the memory function formalism is particularly prone to this unintended scenario, as we discuss later.\\

 As an illustration of a phenomenological model, we consider the case where we model the dynamic structure factor by a sum of three Lorentzian lines, a straightforward extension of the Rayleigh-Brillouin triplet observed in light scattering experiments \cite{Burkel,Lovesey}. The model, when applied to $S_{sym}(q,\omega)$ reads:
\begin{equation}
\pi S_{sym}(q,\omega)= \frac{A_h\Gamma_h}{\omega^2+\Gamma_h^2}+\frac{A_s\Gamma_s}{(\omega-\omega_s)^2+\Gamma_s^2}+\frac{A_s\Gamma_s}{(\omega+\omega_s)^2+\Gamma_s^2}.\label{wrong}
\end{equation}

For brevity, the $q$-dependencies of all variables have been dropped. At first sight this looks like an exact extension of the solutions to the hydrodynamics equations, with the sum $A_h+2 A_s$ equal to $S_{sym}(q)$, as revealed by the direct integration over $\omega$ of Eq. \ref{wrong}. In here, the sound propagation frequency is $\omega_s$ with damping rate $\Gamma_s$. The width of the central line is given by $\Gamma_h$, and the respective amplitudes of these contributions in $S_{sym}(q,\omega)$ are given by $A_s$ and $A_h$, all of which are to be determined through a fitting procedure.\\

However, this model does not have the correct line shapes for the contributions of the propagating modes and it does not satisfy the f-sum rule. As such, it will likely not render the correct propagation frequencies, and it might even give a satisfactory fit to the data when the model should have been rejected. This can be seen by inspecting a 3 x 3 matrix $G(q)$ representing three Lorentzian lines, as discussed in Section \ref{lorentzian}. This model, which has 5 free parameters (four matrix elements, and the prefactor $S_{sym}(q,\omega))$, has amplitudes for the propagating modes that are, in general, complex. Hence, when we fit to a sum of 3 Lorentzian lines, given by the expression below, we do not end up with the type of expression depicted in Eq. \ref{wrong}. In detail, the sum of 3 Lorentzian lines that follows from the projection formalism reads:
\begin{equation}
S_{sym}(q,\omega)=Re [\cfrac{A_h}{i\omega+\Gamma_h}+\cfrac{A_{s,1}}{i\omega+\nu_{s,1}}+\cfrac{A_{s,2}}{i\omega+\nu_{s,2}}],
\label{smurf}
\end{equation}
with, for propagating soundmodes,  $A_{s,1}=A^*_{s,2}$ and
$\nu_{s,1}=\Gamma_s+i\omega_s=\nu^*_{s,2}$. Since the amplitudes of propagating sound modes are complex, taking the real part of Eq. \ref{smurf} leads to asymmetric Lorentzian lines: lines that have additional terms in the numerator proportional to $\omega\Gamma_s/\omega_s$ \cite{bafileoverview}. Ignoring this skewedness leads to erroneous values for $\omega_s$ for cases where the ratio of $\Gamma_s/\omega_s$ is no longer negligible. Therefore, while Eq. \ref{wrong} appears to be such a straightforward extension of hydrodynamics that it almost has to be correct, it employs the wrong line shapes. In addition, it may well overestimate the number of free parameters. The model has five parameters, however, one can be expressed as a combination of the other four ensuring that the f-sum rule can be satisfied.\\

Similar to the model of Eq. \ref{wrong}, the memory function formalism itself is also prone to unintentionally ending up being a phenomenological model. The problem arises when generalizing the memory kernel to situations where kernels, such as the visco-elastic one of Eq. \ref{bla}, no longer give a satisfactory fit to the data. A generalization can be achieved by simply adding a decay channel to $M(q,t)$. Doing so, extended memory functions tend to be written in the following generic form \cite{ScopignoLia,li2000}:
\begin{equation}\label{memgeneral}
M(q,t)=\sum_{i} \Delta_i(q)^2 e^{-t/\tau_i(q)}.
\end{equation}
Unfortunately, this generalization (and others similar to Eq. \ref{memgeneral}) is incorrect since it imposes positive values for each of the decay channels that are characterized by decay times $\tau_i(q)$. We demonstrate in the following why this is incorrect. In addition, Eq. \ref{memgeneral} also assumes that the decay channels are purely dissipative; while this may well be correct for liquids in general, within our current understanding of liquids this remains an assumption (for now).\\

As an example, let us consider two possibilities for extending the visco-elastic model of Fig. \ref{viscoelastic} that have been widely employed in the literature. These two extensions are given by
\begin{equation}
M(q,t)= f_{u\sigma}(q)^2[(1-\alpha)e^{-z_1}+2\alpha
z_2\delta(t)]+f_{uT}(q)^2e^{-z_{T}t}, \label{overeasy2}
\end{equation}
and \begin{equation}
M(q,t)= f_{u\sigma}(q)^2[(1-\alpha)e^{-z_{1}t}+\alpha
e^{-z_2t}]+f_{uT}(q)^2e^{-z_{T}t}. \label{overeasy}
\end{equation}

\begin{figure}
\begin{center}              
 \includegraphics*[viewport=-5 0 555 500,width=70mm,clip]{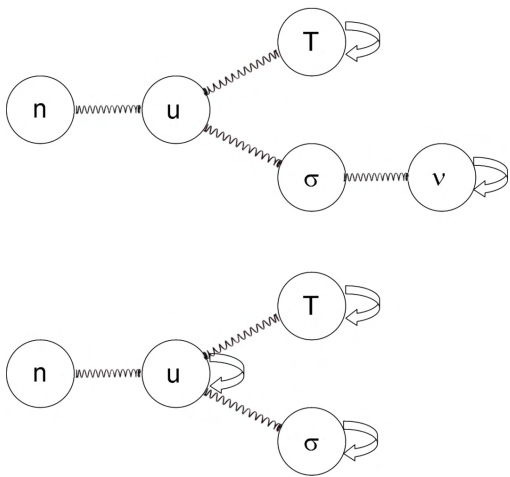}
\end{center}
\caption{The generalization of the decay tree shown in Fig.
\ref{viscoelastic}. The top panel corresponds to Eq. \ref{overeasy}, the
bottom panel corresponds to Eq. \ref{overeasy2}. The symbol '$\nu$'
represents the flux of '$\sigma$'. In the top panel the additional decay
mechanism shows up as a new coupling constant $f_{\sigma \nu}(q)$
and a replacing of the damping rate $z_{\sigma}$ by $z_{\nu}$. In
the bottom the additional mechanism shows up as a new looping arrow at
the microscopic quantity '$u$'.}   
\label{overeasy_modes}
\end{figure}

Both extensions add a decay channel, and both do it in such a way that the overall strength of the channel, $f_{u\sigma}(q)^2$ is unaffected. We show the sketches of the decay tree pertinent to these extensions in Fig. \ref{overeasy_modes}. First, inspecting the extension given in Eq. \ref{overeasy2} (bottom panel Fig. \ref{overeasy_modes}), it is clear that this represents a phenomenological model. Before the modification, the decay tree was exact up to the point where effective damping rates were introduced to capture the very rapid decay of fluctuations in those microscopic variables. Introducing a new decay mechanism is something that is not contained in the Liouville operator, and as such, this extension is an {\it ad hoc} extension.\\

It is of course possible that this {\it ad hoc} extension represents real physics. In particular, in liquid metals we have the conduction electrons that interact with the ionic motion that the decay tree is trying to capture. Therefore, the additional decay mechanism could well represent the electron-ion interaction mechanism. This would be exciting news since there has not been any obvious signature of this interaction mechanism identified in the scattering by liquids \cite{BoveHg,BoveK,ScopignoReview}, only indirect manifestations. However, in the cases where this {\it ad hoc} extension has been used \cite{adhoc}, the additional decay mechanism was interpreted differently, namely as having to do with cage diffusion. The latter interpretation is inconsistent with the exact projection formalism since cage diffusion is already captured in the microscopic variables at the end of the decay tree (Fig. \ref{allmodes}), and thus, the extension of Eq. \ref{overeasy2} is a phenomenological model where we insert a new decay channel without having it being generated through the action of the Liouville operator.\\

The extension given in Eq. \ref{overeasy} and shown in the top panel Fig. \ref{overeasy_modes} is one that the projection formalism does allow for. It has been successfully applied to describe short wavelength excitations in a variety of liquid metals, such as Li \cite{ScopignoLia}, Al \cite{ScopignoAl}, and Na \cite{ScopignoNa}. However, the way it has been applied is inconsistent with the restrictions imposed by the projection formalism, and the results obtained actually point in the direction that this extension fails to capture the dynamics of liquid metals.\\

The problem in how Eq. \ref{overeasy} has been applied in fitting procedures is that $\alpha$, $z_1$, and $z_2$ have all been taken as free parameters. However, as can be seen by inspecting Fig. \ref{overeasy_modes}a, the extension of the decay tree has only introduced one additional parameter compared to the original model (Fig. \ref{viscoelastic}). Thus, only two of the three parameters  $\alpha$, $z_1$, and $z_2$ are independent. We demonstrate this by elimination of the parameter $\alpha$ from the memory function:
\begin{equation}
\label{ubaldo_memory}
M(q,t)=f_{uT}^2e^{-z_Tt}+\cfrac{f_{u\sigma}^2}{z_1-z_2}[z_1e^{-z_2t}-z_2e^{-z_1t}],
\end{equation}
with $z_{1,2}=z_{\nu}/2 \pm \sqrt{z_{\nu}^2/4-f_{\sigma
\nu}^2}$, using the symbols of Fig. \ref{overeasy_modes}. For real $z_{1,2}$, corresponding to the case where
$z_{\nu}/2>f_{\sigma \nu}$ (which is tacitly assumed when fitting
scattering data to the extended memory functions) we see that
one of the exponentials in Eq. \ref{ubaldo_memory} has a negative
amplitude. This finding is in direct contradiction to the generalization of Eq. \ref{memgeneral}. Moreover, upon inspecting the results for the amplitudes reported in the literature \cite{ScopignoReview}, this has never been reported; the values reported for $\alpha$ are between 0 and 1. This strongly suggests that the extension of the visco-elastic model does, in fact, not describe the scattering data in liquid metals.\\

Thus, we are left with a phenomenological model that has more free parameters than allowed for by the projection formalism, that produces results inconsistent with the exact formalism, but that nevertheless describes the scattering data most satisfactorily. This is the danger associated with phenomenological models: they describe the data well, and lead us to interpret the fitting parameters in a certain way. However, when this interpretation is actually inconsistent with itself as shown by a comparison to exact restrictions, then it is time to look beyond such a model (Eqs \ref{overeasy} and \ref{overeasy2}).\\

We note that it might well be the case that Eq. \ref{ubaldo_memory} also manages to describe the data satisfactorily, we do not know until we try. From a physics point of view, an extension of the decay tree of Fig. \ref{allmodes} would make sense if we need to include a very fast process, potentially to do with the interaction between the ions and the electron sea. As mentioned before, the overall effect of the decay of fluctuations in the longitudinal momentum (via a cage diffusion process) yields only a very small contribution to the decay of density fluctuations. It has been estimated to only be an effect of a couple of percentage points even when probing wavelengths as small as the inter atomic spacings  \cite{wouterbook}, and as such it is quite possible that Eq. \ref{ubaldo_memory} would provide a good fit. What is more, scattering experiments are approaching the 1\% degree of accuracy, and we should be using models that incorporate all known restrictions in order to tease out the details of the demise of fluctuations in microscopic variables we do not have direct access to in our scattering experiments.\\

Decay trees that have cross-correlations in them, such as $z_{q\sigma}$ in Fig. \ref{allmodes}, present even more of a challenge. These cross-correlations not only can lead to negative amplitudes in the memory-function and continued-fraction formalisms, but they can even lead to imaginary ones. In other words, even the simple assumption of assuming the amplitudes are real excludes a class of models. As far as we know, there are no theoretical reasons to assume that the cross-correlations equal zero (although computer simulations\cite{gnatzLJ} hint at this possibility).

\section{Details of data analysis}
\label{analysis}
In this section we scrutinize some errors in data analysis methods, errors that can still be found in the recent literature.
\subsection{Symmetrization of scattering data}
We start with a fairly minor probem that has virtually disappeared from our way of doing data analysis. The dynamic structure factor as probed in scattering experiments satisfies the detailed balance condition \cite{vanhove,balu}
\begin{equation}\label{detbal}
S(q,-\omega)=e^{-\beta \hbar \omega}S(q,\omega).
\end{equation}
The projection formalism expresses the dynamics of liquids in terms of static quantities pertinent to a symmetrized version of the dynamic structure factor (see Eq. \ref{main}). Also, the dynamic structure factor that can be obtained from computer simulations is symmetric in energy. There exist infinitely many ways in which the detailed balance condition of Eq. \ref{detbal}      can be utilized to obtain an symmetrized dynamic structure factor with the aim of directly comparing scattering experiments to computer simulations. One of those ways is simply to multiply the measured $S(q,\omega)$ by $e^{-\beta\hbar\omega/2}$. This method was employed to analyze the scattering by inert gases \cite{gnatzAr,vanwellAr,vanwellNe,proptemp,he40k}, but it ran into problems when colder liquids, such as helium, were being investigated \cite{wouterbook}. We stress that the symmetrization given in Eq. \ref{main} is the only way in which the projection formalism can be applied without incurring additional frequency factors; all other symmetrizations would not result in a function that can be modeled according to Eq. \ref{main} and would therefore not yield the poles of the dynamic susceptibility.\\

Of course, we can still symmetrize the data in any way we wish, as long as we include additional frequency factors in our model. With few exceptions \cite{chen}, data are now symmetrized according to Eq. \ref{main}. An example of how results can be misinterpreted when an inappropriate symmetrization is being used can be found in Montfrooij {\it et al.} \cite{proptemp} where scattering data on supercritical helium at 13 K were symmetrized using the factor  $e^{-\beta\hbar\omega/2}$. Following this symmetrization, it was found that the dynamics of this particular liquid state differed greatly from other liquids investigated prior to this date in the sense that the liquid appeared capable of sustaining propagating temperature fluctuations. However, we believe that these results are in doubt given the usage of an incorrect symmetrization procedure. As is evident from Eq. \ref{main}, failure to include the appropriate frequency factor in the model fit can easily lead to distorted lineshapes in models. The bottom line is that only the symmetrization given in Eq. \ref{main} yields a function that can be modeled by any of the projection formalism based methods in such a way that the propagation frequencies and relaxation rates obtained are directly identifiable as the poles of the dynamic susceptibility.

\subsection{Identifying peak positions with decay channels}
In the low-$q$ hydrodynamic regime, the dynamic structure factor is a combination of three modes, the so-called Rayleigh-Brillouin triplet \cite{balu}. Upon probing the liquid on shorter and shorter length scales, the widths (in energy) of the peaks rapidly increase, and the three distinct peaks observed in light scattering become less well defined \cite{bafilehydro}, and can even merge into a fairly featureless $S(q,\omega)$ \cite{gnatzAr}. In order to infer the details from the merged features, it is necessary to resort to model fitting. However, model fitting is time consuming, and it comes with a certain degree of subjectivity. In order to bypass the fitting procedure, it used to be common practice to inspect the longitudinal current-current correlation function instead. This function is directly accessible to scattering experiments as it is given \cite{balu} by $\omega^2S(q,\omega)$.\\

The idea behind inspecting this function is that propagating modes will become more pronounced, whereas non-propagating modes centered around $\omega$= 0 will be suppressed. This way, the peak positions in  $\omega^2S(q,\omega)$ can be directly read off from the data, independent of any model fitting. These peak positions are then identified with the propagation frequencies of the modes of interest, such as short-wavelength sound modes.\\

\begin{figure}       
\begin{center}              
\includegraphics*[viewport=50 100 550 400,width=80mm,clip]{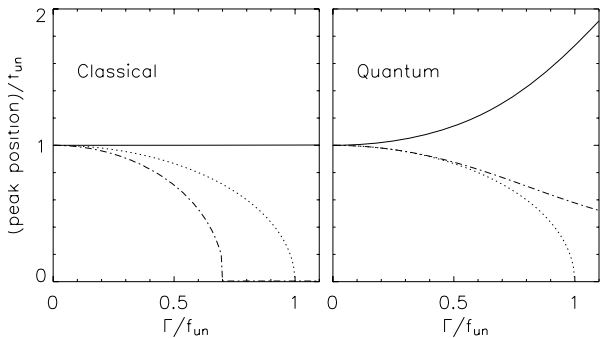}
\end{center}
\caption{The peak positions of $S(q,\omega)$ (dashed-dotted line),
those of the longitudinal current-current correlation function $\omega^2S(q,\omega)$ (solid line)
and the poles of the dynamic susceptibility (dotted line,
corresponding to the sound propagation frequency $\omega_s =
\sqrt{f_{un}^2-\Gamma^2}$). This figure was calculated for the case where the excitations are
described by a damped harmonic oscillator model. The figure on the left
corresponds to very high temperature, the figure on the right
corresponds to $T$= 0 K. Note that the positions of the peaks in neither
$S(q,\omega)$ nor in $\omega^2S(q,\omega)$ are a good measure of the frequency
of the underlying excitation when $\Gamma >$ 0.1$f_{un}$.}
\label{peak_positions2}
\end{figure}

Unfortunately, the idea behind this procedure is only valid in the case where the excitations are almost elementary excitations, that is, excitations whose decay rates are much smaller than the propagation frequencies that are being determined. We illustrate this in Fig. \ref{peak_positions2} for the easiest possible case where the dynamic structure factor consists only of two propagating sound modes. As can be seen from this figure, the peak positions in  $\omega^2S(q,\omega)$ do not correlate in any predictable manner with the poles of the dynamic susceptibility once the decay rates exceed about 10\% of the propagation frequency. For more complicated models, such as those that contain one pair of propagating modes and one or two central lines, identical considerations hold. As such, one should not attempt this shortcut, but instead perform a full model fit in order to extract the information of interest, namely the poles of the dynamic susceptibility.

\subsection{Imposing too restrictive conditions}\label{restrictive}
When discussing phenomenological models we pointed out how we sometimes leave too many parameters as adjustable parameters, resulting in the potential of accepting models that should have been rejected. The opposite also occurs when we impose conditions on our models that are too restrictive in the sense that we implicitly impose a certain outcome on our models in instances where liquids can also display a different outcome. One such restriction we already discussed when we looked at the description of the scattering data by a sum of Lorentzian lines is that all had real amplitudes; a restriction that is not met by propagating modes. The restriction we discuss in this section is that of imposing propagating modes on a system while there is no reason a priori for this system to sustain propagating modes. This practice has a long and checkered history \cite{Fak}, and it is a practice that is not restricted to liquids. For instance, recent data \cite{magnon} on magnetic scattering by high temperature superconductors have been analyzed with this restriction imposed upon the model, leading to conclusions not fully supported \cite{prbw} by the data.\\

We illustrate the above by looking at a damped harmonic oscillator. Every physics student is familiar with the solutions to this model, yielding the damped, critically damped and overdamped solutions depending on the ratio of the damping rate compared to the undamped frequency. When such an oscillator is probed by means of scattering experiments, we find two peaks in the dynamic structure factor for the case of the damped harmonic oscillator. These peaks correspond to exciting an oscillator quantum, and to the absorption thereof. The detailed balance condition of scattering experiments ensures that the strength of the excitation peak is always greater than the strength of the absorption peak, with the latter being absent at the very lowest temperatures probed in experiments on liquid helium.\\

In the case of overdamped oscillators, the solutions (in time) are given by exponentially decaying signals with two different decay constants. In scattering experiments this corresponds to two central (in energy) Lorentzian lines, one with a positive amplitude, and one with a negative amplitude. Both scenarios are captured by the following model where $f_{un}$ plays the role of the undamped frequency, and $z_u$ that of the damping rate (which we can loosely think of as the Full Width at Half Maximum in a scattering experiment).
\begin{equation}
S_{sym}(q,\nu ) =\cfrac{S_{sym}(q)}{\pi
}  \cfrac{f_{un}^2 z_u} {(f_{un}^2-\omega ^2)^2+(\omega z_u)^2}.
\label{dhoEq.}
\end{equation}

Note that this is a special case of Eq. \ref{failsafe} with B= 0 following from the application of the f-sum rule. For the case of propagating modes, with propagation frequency $\omega_s=\sqrt{f_{un}^2-(z_u/2)^2}$ and decay rate $\Gamma_s=z_u/2$, this equation can be rewritten as the sum of two Lorentzian lines as follows:
\begin{equation}
\begin{aligned}
S_{sym}(q,\omega)=\cfrac{S_{sym}(q)}{2\pi} \times \\[10pt]
\left (
\cfrac{2\Gamma_s(q)+\omega\Gamma_s(q)/\omega_s(q)}{\Gamma_s(q)^2+[\omega+\omega_s(q)]^2}
+\cfrac{2\Gamma_s(q)-\omega\Gamma_s(q)/\omega_s(q)}{\Gamma_s(q)^2+[\omega-\omega_s(q)]^2}
\right ).
\end{aligned}
 \label{dho_asym}
\end{equation}
Note that each of the two Lorentzian lines are asymmetric in energy, a result of the fact that the amplitudes of propagating modes are complex. Fitting the data to Eq. \ref{dho_asym} will yield the poles of the dynamic susceptibility for the case where the scattering is described by a damped harmonic oscillator. For cases where the scattering includes more than two Lorentzian lines, the expression to fit to will be more complex, but the essence remains unchanged in the sense that propagating modes are represented by a pair of Lorentzian lines located at $\pm \omega_s(q)$ in $S_{sym}(q,\omega)$.\\

The problem with using Eq. \ref{dho_asym} from the outset \cite{Fak} rather than fitting to the general model of Eq. \ref{dhoEq.} is that it does not allow for overdamped modes. This can be seen by inspection of Eq. \ref{dho_asym} in the limit $\omega_s \rightarrow 0$, but it is also immediately clear that this equation with $\omega_s = 0$ can never describe two non-propagating (overdamped) modes with different damping rates [given by $\Gamma_{1,2}=z_u/2\pm \sqrt{z_u^2/4-f_{un}^2}$]. The fitting procedure will be very deceptive, however. The fitting routine will find a minimum corresponding to a finite propagation frequency, and the agreement between the model and the data will look reasonably good.\\

\begin{figure}
\begin{center}              
\includegraphics*[viewport=00 -20 515 320,width=95mm,clip]{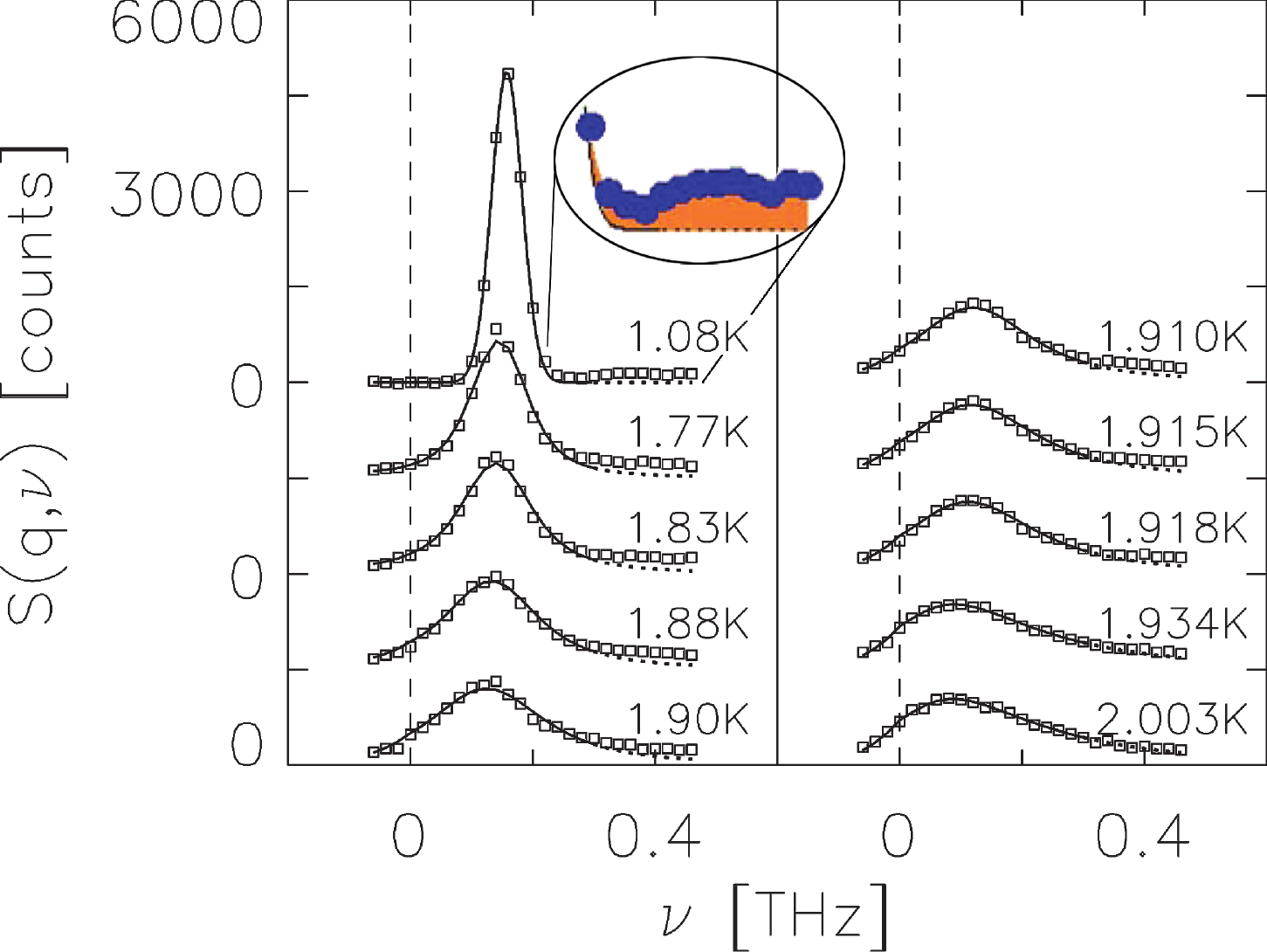}
\includegraphics*[viewport=100 370 530 690,width=80mm,clip]{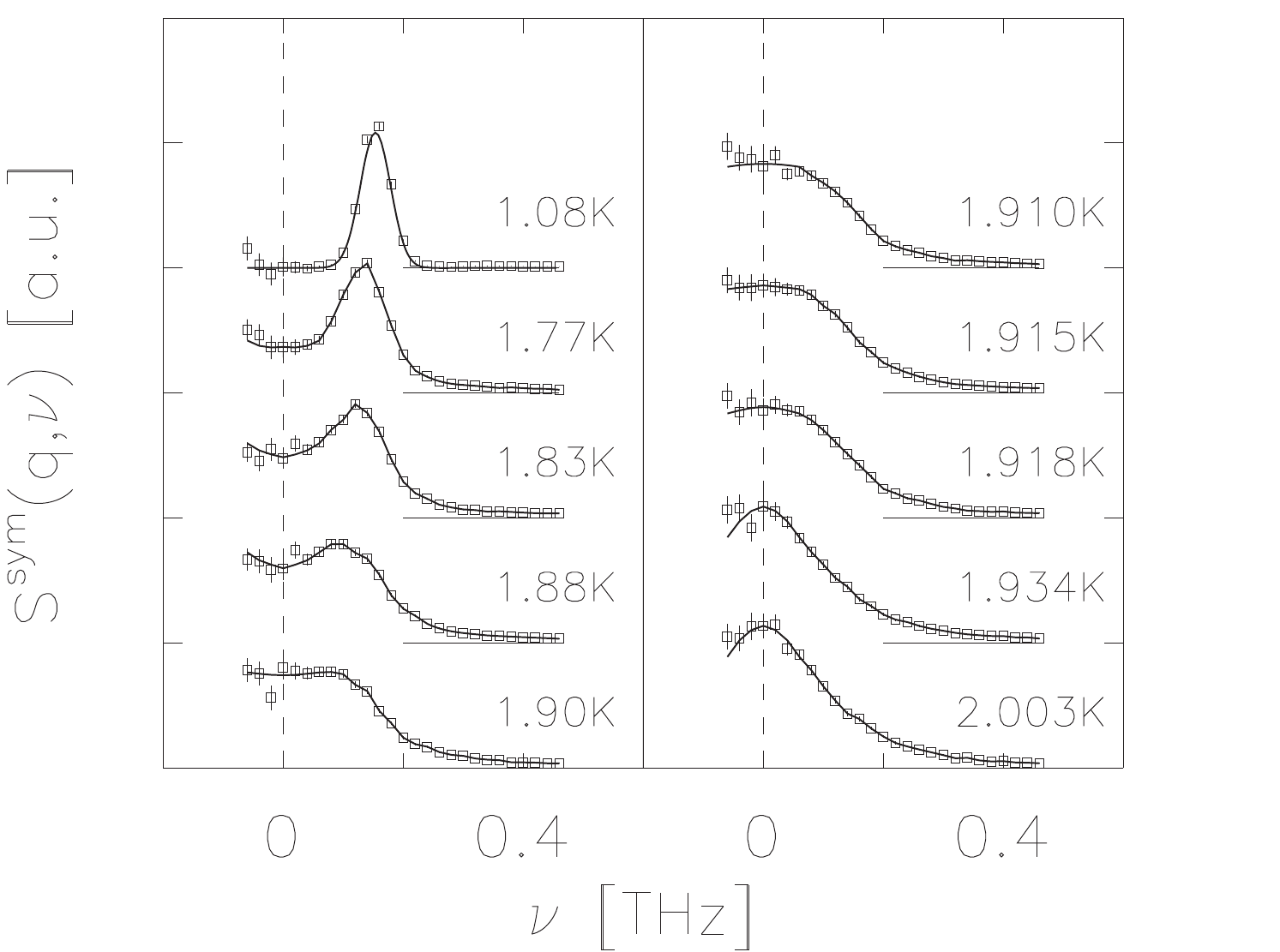}
\end{center}
\caption{Figure reproduced with permission from ref. [\onlinecite{Svenssonsoftrot}]. The top panel shows the dynamic structure factor $S(q,\omega= 2\pi \nu)$ at constant $q$ for liquid helium under pressure (constant density of 0.1715 g/cm$^{3}$) for temperatures indicated in the graph. The superfluid transition for this density is at T= 1.9202 K. The data show a broadening of the excitations when the temperature is raised, with very few changes apparent upon crossing through the normal fluid to superfluid transition. The mode appears to be propagating at either side of the transition. The bottom panel displays the identical data, but now the trivial frequency factor of Eq. \ref{main} has been removed. It can now be seen that the mode which is propagating at the lowest temperatures broadens and softens upon raising the temperature, with the mode becoming non-propagating once the normal fluid phase has been reached. Line shape distortions that appear to yield propagating modes in the case of overdamped modes are to be expected when $\beta \hbar \omega \gg$ 1.}   
\label{helium}
\end{figure}

The origin of using Eq. \ref{dho_asym}, as opposed to using Eq. \ref{dhoEq.} (which does not impose propagating modes from the outset) can be traced back to the frequency factor in Eq. \ref{main}. Because of this factor, when $\beta \hbar \omega \gg 1$, even non-propagating modes will peak at finite frequencies in $S(q,\omega)$. We illustrate this behavior in Fig. \ref{helium}. This situation occurs in quantum fluids such as liquid helium, but also at high temperatures in solids when the magnetic dynamics are probed at very high frequencies \cite{magnon,prbw}. Cursory inspection of the dynamic structure factor in these cases (see Fig. \ref{helium}) appears to be convincing evidence for propagating modes, and hence the use of Eq. \ref{dho_asym} appears justified when, in fact, it is not. While this is an easy mistake to avoid, it is a mistake that has seen many instances in the literature \cite{prbw}. A discussion pertinent to $^3$He is given in references [\onlinecite{albergamohe}] and [\onlinecite{Schmets}].\\

In addition, while Eq. \ref{dhoEq.} identifies the propagation frequencies as  $\omega_s=\sqrt{f_{un}^2-\Gamma_s^2}$, quite often the undamped frequencies $f_{un}$ are reported \cite{BoveK,BoveGa,HosokawaSi,CazzatoNi,albergamohe} as being the frequencies of interest. However, given that the poles of the dynamic susceptibility (in the case of propagating modes) are given by $i\Gamma_s \pm \sqrt{f_{un}^2-\Gamma_s^2}$, presenting the results for  $\omega_s$ and $\Gamma_s$, together with their dependencies on momentum transfer, temperature and density, is a better practice.

\subsection{Comparison to computer simulations}
When we perform molecular dynamics computer simulations, we calculate the intermediate scattering function whose Fourier transform yields the dynamic structure factor, and sometimes we also determine other intermediate scattering functions, such as the one describing the correlations between the microscopic density and microscopic energy density. In all, there exist three independent correlation functions \cite{gnatzLJ}. All these correlation functions can be modeled within the projection formalism by the matrix $G(q)$ describing all correlations between all microscopic variables. In fact, computer simulations are very helpful because all the off-diagonal elements of this matrix, namely the coupling constants representing equal time correlation functions, follow directly from the simulations without having to resort to a fitting procedure. Thus, models describing the time-dependence of the correlation functions as determined by molecular dynamics computer simulations only have the decay rates (diagonal elements) as free parameters.\\

However, there is an important difference in the correlation functions determined in computer simulations, and those measured in scattering experiments. In real liquids, the intermediate scattering function is a complex function, related to the asymmetry (in energy) of $S(q,\omega)$. In computer simulations, the intermediate scattering function takes on strictly real values, and its Fourier transform yields a function that is symmetric in energy. This statement of course merely reflects the fact that we cannot use computers to solve classical equations of motion and hope to reproduce a system obeying quantum mechanical equations. In order to deal with this in practice, we compare the results of computer simulations to the symmetrized dynamic structure factor given by
\begin{equation}
S_{sym}(q,\omega)=\cfrac{1-e^{-\beta \hbar \omega}}{\beta
\hbar \omega} S(q,\omega).\label{sym1}
\end{equation}

It is clear that in the classical, high-temperature limit ($\beta \hbar \omega \ll$ 1) the symmetrized structure factor can be directly compared to computer simulations. Nonetheless, even when using the symmetrized function, essential differences remain between the true dynamic structure factor, and the one that is calculated in computer simulations. This is best illustrated by scrutinizing the equal time correlation functions that are accessed in scattering experiments, such as $S_{sym}(q)$ and $f_{un}(q)$. \\

Integrating Eq. \ref{sym1} over $\omega$ on both sides of the equation immediately demonstrates that $S_{sym}(q)$ is not related to the zeroth frequency moment of $ S(q,\omega)$ (being the the static structure factor $S(q)$), but rather to the static susceptibility $\chi(q)$ (the (-1)$^{st}$ frequency moment). In detail: \cite{montfrooijEE}
\begin{equation}
\int_{-\infty}^{\infty} \omega ^{-1} S(q,\omega ) d\omega= \hbar \pi
\chi(q)= \hbar \beta S_{sym}(q)/2. \label{sumrule1}
\end{equation}

Similarly, while in computer simulations $f_{un}$ is given by the ratio of the second and zeroth frequency moments, in real fluids (atomic mass $m$) it is given by the ratio of the first and (-1)$^{st}$ frequency moments and reads:
      \begin{equation}
  f_{un}(q) = \cfrac{q}{\sqrt{S_{sym}(q)\beta m}} =
 \cfrac{q}{\sqrt{2\pi m\chi(q)}}.
  \label{II8}
  \end{equation}

Thus, computer simulations and scattering experiments access different quantities. In simulations, the dynamics is given by even frequency moments, in real liquids the odd frequency moments play the leading role. This statement holds true independent of the temperature of the actual system. While this may be surprising, the apparent controversy is removed once we realize that the classical equivalent of the dynamic structure factor for real liquids does not exist in nature; it only exists in simulations.\\

We can best appreciate the difference between computer simulations and a real liquid by looking at a pure quantum liquid, namely superfluid helium. The dynamic structure factor of liquid helium does not change appreciably between 0 K and 1 K, with sharp, elementary excitations clearly visible at frequencies independent of temperature. In addition, the static structure factor of helium also remains virtually unchanged. Inspecting Eq. \ref{II8}, we can see that the reason behind this is that the static susceptibility $\chi(q)$ is temperature independent at low temperatures. Had we equated $S_{sym}(q)$ with the static structure factor $S(q)$ instead, then we would have found that $f_{un}(q)$ would depend strongly on temperature, something not borne out by experiments. Conversely, at very high temperatures, we find that $S_{sym}(q)$ and $S(q)$ are almost indistinguishable while $\chi(q)$ is inversely proportional to the temperature.\\

Given this, we must be careful in how we compare experiments and simulations. For one, the quantity  $S_{sym}(q)$  we derive from model fitting to our data is not the static structure factor $S(q)$. Often, however, this quantity $S_{sym}(q)$  is presented \cite{never} in the literature as being $S(q)$ rather than the static susceptibility $\chi(q)$. If we wish to show $S(q)$ based on our scattering experiments, then we either have to carry out a numerical integration over our measured dynamic structure factor, or alternatively, we have to take our model, divide it by the frequency dependence shown in Eq. \ref{sym1} and carry out the integration numerically. Note that this integration cannot be done analytically. As far as we are aware, the second method has never been applied in the literature. Instead  $S_{sym}(q)$  is identified \cite{never} as being $S(q)$.\\

Similar considerations hold for $ f_{un}(q)$ as determined from scattering experiments. Sometimes this coupling constant is used to indirectly determine the static structure factor, however, it can only yield the static susceptibility. This misidentification might even be the reason that we tend to determine  $ f_{un}(q)$ from experiments through a fitting procedure, even though in principle it should not be a free parameter. Since this parameter is exactly given by the ratio of the first frequency moment with the static susceptibility, rather than the ratio of the second frequency moment with the static structure factor, we have to acknowledge that we do not have a good determination of the static suscpetibility. In order to determine $\chi_q$  from our scattering data we have to divide the dynamic structure factor by energy, and integrate over all energy transfers. But this procedure is not very accurate around $E$= 0 because of spurious scattering from the sample environment. In other words, it cannot be determined with a very high degree of accuracy from experiments. Note also that it does not follow from the overall prefactor $S_{sym}(q,\omega)$ in an unambiguous manner since this prefactor also needs to account for the overall scale factor that incorporates the neutron/x-ray flux and monitor efficiency.\\

\begin{figure}
\begin{center}
\includegraphics*[viewport=150 0 800 500,width=80mm,clip]{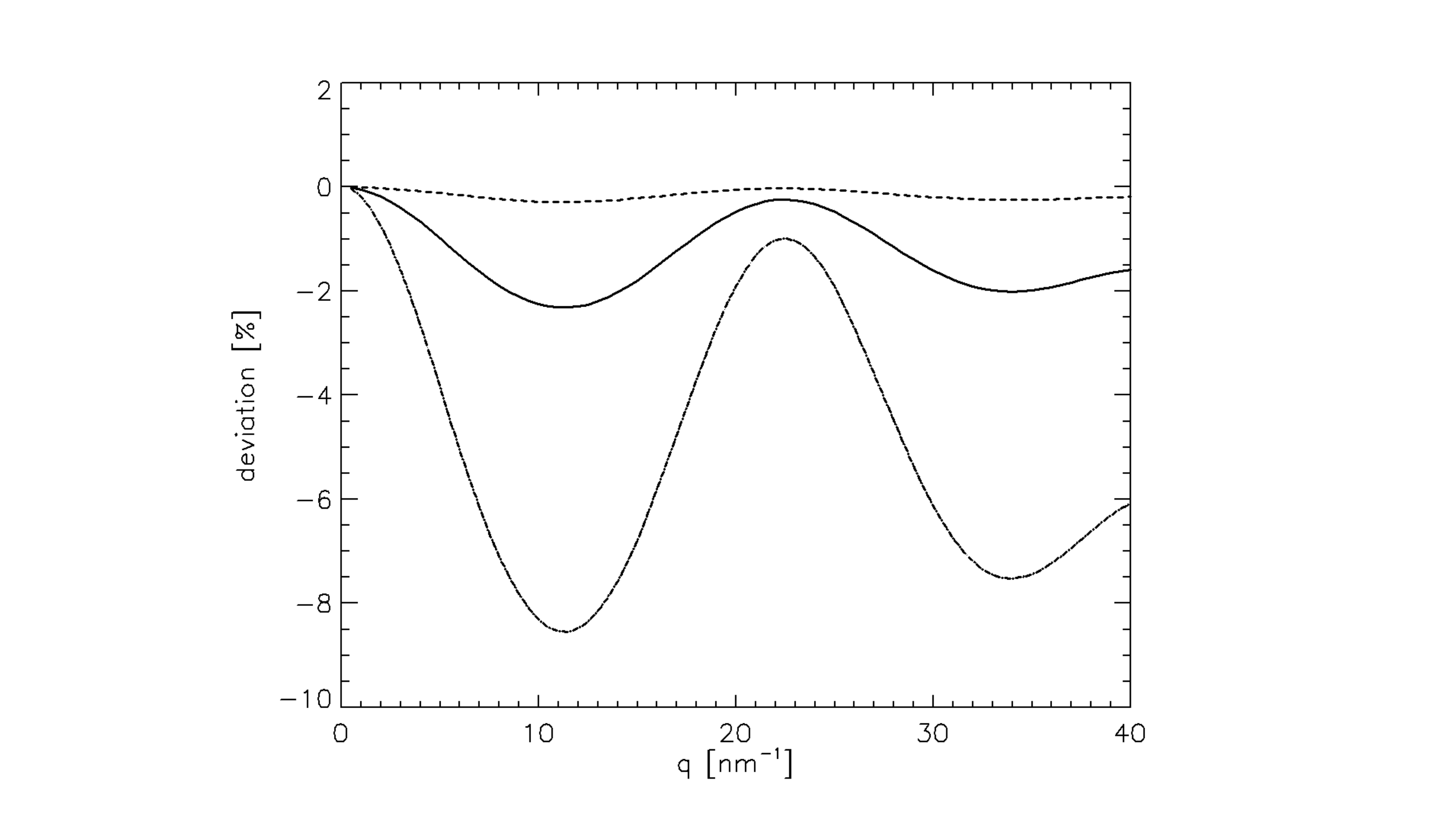}
\end{center}
\caption{When identifying the prefactor $S_{sym}(q)$ of Eq. \ref{main} obtained from model fitting to experiments with the static structure factor $S(q)$, errors are being introduced. The lines in the figure show the deviations of the ratio $S_{sym}(q)/S(q)$ from unity, the idealized case. In order to arrive at these curves, we used a parameterized visco-elastic model that roughly reproduced the observed frequencies and linewidths for the propagating modes in the systems identified below. All calculations were performed using a static structure factor peaked at $q$= 22 nm$^{-1}$. We carried out a numerical integration of the model in order to obtain $S(q)$. The top line has been calculated for Ga at 325 K, the middle line for Al at 1,000 K and the bottom curve for Neon at 40 K. Note that the largest deviations are expected for $q$-values near the maximum of the dispersion curve.} \label{donotdothis}
\end{figure}

Irrespective of how accurately we can determine $S_{sym}(q,\omega)$ directly from our data, there is no merit in assuming that  $S_{sym}(q)$ is identical to the static structure factor: the latter can only be determined by numerical integration, or by having a detector carry out the energy integration for us. We show the scale of errors made in identifying $S_{sym}(q)$ with $S(q)$ in Fig. \ref{donotdothis}.\\

\section{Modeling in the absence of detailed knowledge of the Liouville operator} \label{transverse}
When the Liouville operator is known, as is the case for most liquids, then we can let it act on successive microscopic variables and populate the matrix with the relevant couplings and decay rates. This is the procedure we have discussed thus far for the exact formalisms, the procedure that automatically ensures we have the most restrictive model to fit our data with. However, what if we find a new excitation in a liquid that cannot be accommodated within this scheme? This possibility became reality in 2009 when Hosokawa and co-workers reported \cite{hosokawa} on the existence of transverse acoustic excitations in Liquid Ga; in addition, in 2010, Giordano and Monaco inferred transverse excitations in liquid sodium \cite{travmodes}.\\

\begin{figure}
\begin{center}
\includegraphics*[viewport=40 100 550 730,width=80mm,clip]{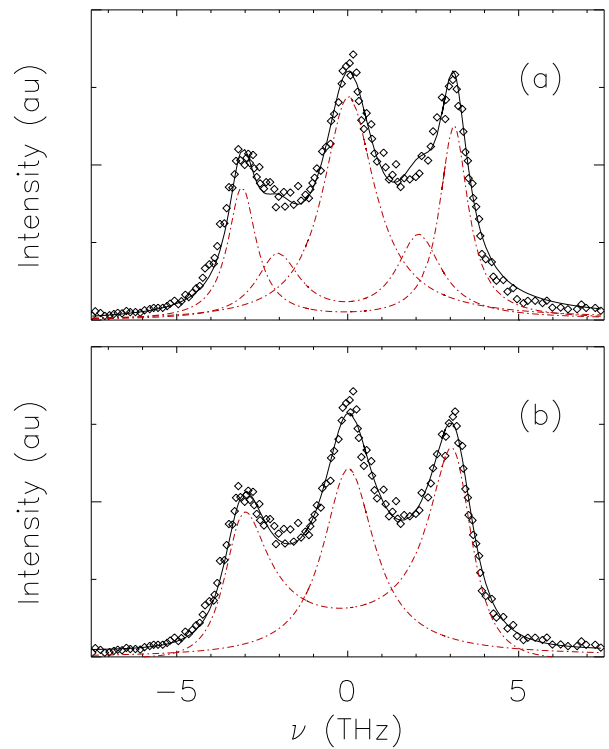}
\end{center}
\caption{Panel (a) shows x-ray scattering data on liquid sodium at 393 K at $q$= 7.6 nm$^{-1}$ as a function of frequency $\nu= \omega/2\pi$. The data have been reproduced from reference [\onlinecite{travmodes}]. The solid line through the data points is the result of a fit to five Lorentzian lines that do not take any asymmetry into account. The individual lines are shown by the dotted curves. Giordano and Monaco  \cite{travmodes} interpreted the two pairs of propagating modes derived from such a fit as longitudinal and transverse excitations. Note, we did not reproduce the errorbars in this plot in order to allow for a better visual comparison between data and model. Panel (b) shows the fit to the (extended) hydrodynamic model (see Table I and Eq. \ref{smurf}) that takes the asymmetry of propagating modes into account. This fit is equivalent to using Eq. \ref{barocchi} with three Lorentzian lines. The quality of the fit in panel (b) is actually slightly better than the one in panel (a). As such, the data at this q-value do not show direct evidence for the presence of transverse modes.} \label{sodium}
\end{figure}

When analyzing the x-ray scattering spectra of liquid sodium, Giordano and Monaco discovered that while the spectra at the lowest momentum transfers (up to 3 nm$^{-1}$) could be described by the sum of three Lorentzian lines, the spectra for larger $q$ required 5 Lorentzian lines: a central line and two pairs of propagating excitations. Moreover, the excitation frequencies of one pair of propagating modes were very similar to the propagation frequencies of transverse excitations in the solid phase. The authors concluded that liquid sodium is capable of sustaining propagating transverse excitations. This came as a surprise since liquids are believed to be incapable of sustaining propagating transverse modes. We reproduce a typical example of the scattering data in Fig. \ref{sodium}a.\\

Since this discovery, the presence of transverse modes has been inferred from the scattering data on other liquids, and computer simulations have been performed to develop a better understanding of the mechanism behind these modes. The problem behind understanding the origin of these transverse modes is that the Liouville operator that is used for liquids does not have a term in it that couples longitudinal to transverse excitations. The standard terms of the Liouville operator are proportional to $\sim \cdot \partial/\partial \vec{r}_i$ and to  $\sim \cdot \partial/\partial \vec{v}_i$, with the index 'i' running over the particles in the liquid at position $\vec{r}_i$ and with velocity $\vec{v}_i$. The dot in front of both operators indicates that we take the scalar product. Because of this, all new microscopic variables that are being generated through the action of the Liouville operator are longitudinal variables, not transverse ones. In order for there to be coupling to transverse microscopic variables, we need a term in the Liouville operator that generates this coupling, such as a rotational term. However, it is unclear where such a rotational term should come from in a mono-atomic liquid such as sodium.\\

In the absence of knowledge of the full Liouville operator, the projection formalism cannot generate exact couplings, and as such, we cannot impose all the restrictions on our fitting models that we can for liquids that only support longitudinal excitations. The question now is one of what is the most restrictive model that we can impose on our data? Before we address this question, we first  look at a similar situation that was encountered in mixtures, followed by a closer scrutiny of the fitting model employed to reach the conclusion on the existence of transverse modes in sodium.\\

Bosse {\it et al.} discovered \cite{bosse} by doing molecular dynamics computer simulations that binary mixtures are capable of sustaining propagating excitations that travel at speeds greatly exceeding the hydrodynamic sound velocity. They dubbed these excitations fast sound. Their existence was verified \cite{fastexp} through neutron scattering experiments. It was only after these simulations and experiments that the appropriate projection formalism based on the Liouville operator for mixtures was developed \cite{mixtures}. Since then, slow sound modes \cite{slowmodes} have been observed as well, and the theory of the dynamics in liquid mixtures is now on a solid foundation \cite{bryk}. This paragraph illustrates that the interplay between simulations, theory and experiment can be successful in first identifying new excitations in the absence of complete knowledge of the Liouville operator.\\

Before we can be certain about the existence of new excitations (Fig. \ref{sodium}a), we must make sure that the necessity to include more Lorentzian lines is not due to having used a model that is inappropriate for the system under study. We show what we mean by this by displaying a fit to the sodium data using only 3 Lorentzian lines, but with the asymmetric contributions taken into account. We show the results in Fig. \ref{sodium}b. As can be seen from the agreement between the x-ray data and the model fit, there is no immediate need to resort to 5 Lorentzian lines at this particular wave number. From comparison between the two panels of Fig. \ref{sodium} it would appear that the Lorentzians employed in the original fit were restricted to be symmetric around the excitation energies rather than having a pronounced asymmetry, resulting in the need to employ more than 3 Lorentzian lines in order to compensate for the excluded asymmetry.\\

When we try to fit the scattering data at higher momentum transfer, we also find that three lines no longer suffice. However, if we try a decay tree such as the one shown in Fig. \ref{viscoelastic}, we find that it is still possible to obtain a good fit without having to resort to additional propagating modes. As such, in order to conclude that the presence of transverse modes is borne out by the data, we would actually have to scrutinize the fitting parameters that are obtained as a function of momentum transfer, and look for any discontinuities and/or stand-out behaviors as a function of probing wavelength. We have not done this for two reasons. First, our objective is not to deduce whether transverse modes are present in any particular liquid or not, our objective is to ensure that all data are analyzed using the appropriate models. In this case, it would appear that using asymmetric Lorentzians leads to a different conclusion than employing symmetric Lorentzians. Since there is no theoretical underpinning for using the latter, we urge to drop their usage from analyzing scattering data. Second, the inference of transverse modes is not based solely on scattering experiments; numerous computer simulations \cite{sim1,sim2,guarini,sim3,guarini2} have been performed that are consistent with the presence of such modes.\\

As we mentioned earlier, the model employed by Hosokawa {\it et al.} \cite{hosokawa} that did not produce a good fit to their data (one central Lorentzian plus a damped harmonic oscillator) is a model that does not satisfy the f-sum rule. The f-sum rule on $S(q,\omega)$ corresponds to the second frequency moment of $S_{sym}(q,\omega)$. This moment diverges unless the central Lorentzian is compensated for by the high-energy contribution of the B term: C$\Gamma$=-B (Eq. \ref{dothis}).  As such, it should not come as a surprise that it does not fit well to highly accurate scattering data. Their improved model \cite{hosokawa,hosokawa2}, consisting of adding another damped harmonic oscillator to the fit function does not satisfy the f-sum rule either: choosing all the B$_i$ in Eq. \ref{dothis} to be equal either implies the absence of a central Lorentzian, or a failure of the model to satisfy the exact f-sum rule. A better approach would have been to use Eq. \ref{failsafe}, or one of the many exact equivalent representations. Should such an exact model still have failed, then the experimental evidence for the existence of transverse modes would have been (even) stronger. We note that the approach by Zanatta {\it et al.}\cite{zanatta} is very similar to the procedure of  Hosokawa {\it et al.} with the difference in fitting procedure being that the two damped harmonic oscillators are coupled; however, the central Lorentzian is not encorporated into the sum rules.\\

When faced with (the possibility of) a new excitation, what is the best approach? Barocchi and Bafile \cite{barocchi}  have shown that any correlation function can always be expressed as the sum of Lorentzians. In this summation, the non-propagating modes are symmetric around $\omega$= 0 in $S_{sym}(q,\omega)$, while the non-propagating modes show up in pairs as asymmetric Lorentzians. In fact, their \cite{barocchi}  approach holds for any correlation function, including self-correlation functions or correlations between microscopic density and temperature. Thus, any correlation function $C_{sym}(q,\omega)$, symmetrized in the same way as $S_{sym}(q,\omega)$, can be written as (dropping the q-dependence for brevity):

\begin{equation}
\begin{aligned}
C_{sym}(q,\omega)= \cfrac{1}{\pi} \sum_r \cfrac{A_r \Gamma_r}{\omega^2+\Gamma_r^2}+ \\[10pt]
 \cfrac{1}{\pi}\sum_c \cfrac{A_c' \Gamma_c-A_c''(\omega-\omega_c)}{(\omega-\omega_c)^2+\Gamma_c^2}+\cfrac{A_c' \Gamma_c+A_c''(q)(\omega+\omega_c)}{(\omega+\omega_c)^2+\Gamma_c^2}
\end{aligned}
\label{barocchi}
\end{equation}
In here, the two summations run over all the non-propagating and propagating modes. The amplitudes of the propagating modes  at $\omega=\pm\omega_c(q)+i\Gamma_c(q)$ are complex ($A_c'(q)+iA_c''(q)$), the amplitudes $A_r(q)$ of the non-propagating modes at $\omega=i\Gamma_r(q)$ are real. Inspecting the above equation we see that every non-propagating Lorentzian comes with two unknown parameters (the real damping rate and the real amplitude), and every pair of propagating modes comes with 4 unknown parameters (the propagation frequency, the damping rate, and the real and complex parts of the amplitude). In the absence of full knowledge of the Liouville operator, we can no longer relate the amplitude to the propagation frequency and damping rate as we were able to do using the projection formalism.\\

The number of unknown parameters is reduced by the sum rules as discussed in previous section, such as the zeroth sum rule that yields the static factor $C(q)= \sum_{i=r,c} A_i(q)= \int C_{sym}(q,\omega)d \omega$ as well as by demanding the existence of higher moments. Note that the summation in Eq. \ref{barocchi} has been separated in propagating and non-propagating modes. Needless to say, it should not be assumed prior to doing a fitting procedure that a mode is propagating or not as this might well lead to identifying a non-propagating mode as being a propagating one. Also note that the propagating modes are represented by asymmetric Lorentzians, an asymmetry which is easy to identify in Fig. \ref{sodium}: fitting Eq. \ref{barocchi} to the data shown in Fig. \ref{sodium}a we find that they are well described by the sum of three Lorentzian lines.\\

In order to avoid these difficulties (and arrive at Eq. \ref{failsafe}), we note that it is possible to fit to a sum of Lorentzian lines and determine the propagation frequencies and damping rates without imposing any prior assumptions on the model, such as that propagating modes should exist. Any two Lorentzian lines can be added together to yield the following expression:

\begin{equation}
 \cfrac{1}{\pi} \cfrac{A+B\omega^2}{(\omega^2-\Omega^2)^2+z^2\omega^2}.
\label{anytwo}
\end{equation}
In this equation, $A$, $B$, $\Omega$, and $z$ are real fitting parameters. The desired frequencies of the modes are determined by $\Omega$ and $z$: we have a pair of propagating modes if $\Omega > z/2$ (with damping rate of $z/2$ and propagation frequencies of $\sqrt{\Omega^2-z^2/4}$), and we have two non-propagating modes if   $\Omega < z/2$ (with two different damping rates $\Gamma_1$ and $\Gamma_2$ given by $\Omega=\Gamma_1\Gamma_2$ and $z$ = $\Gamma_1 + \Gamma_2$). Only after the fitting procedure will it become known if we are looking at propagating modes, or non-propagating. This form captures both possibilities for non-propagating modes, namely that we have two overdamped modes, or that we simply have two damped modes that are not-connected to each other. The fit parameters $A$ and $B$ would distinguish between these two cases.\\

\begin{figure*}
\begin{center}
\includegraphics*[viewport=15 100 540 650,width=100mm,clip]{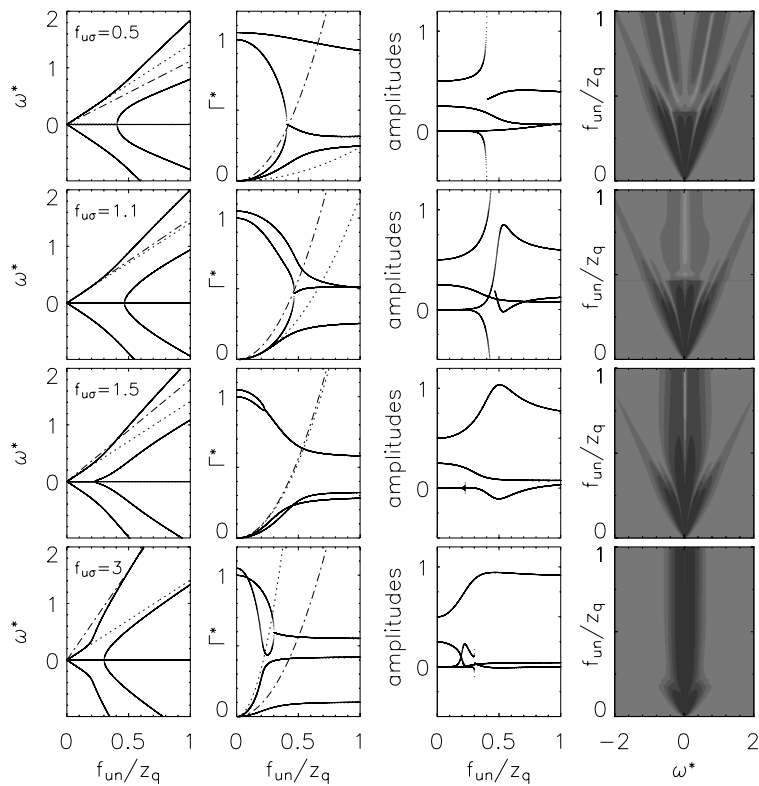}
\end{center}
\caption{Figure reproduced from Ref. \cite{wouterbook}. An illustration of the dependence of the amplitudes, damping rates and propagation frequencies on the coupling parameters between the microscopic variables. The plots shown were calculated for the 5x5 matrix shown in Eq. \ref{m55} and depicted in Fig. \ref{allmodes}. The main takeaway from this plot is that even when the matrix elements vary gradually, the ensuing modes (Lorentzian lines) vary rapidly, both in amplitude and propagation frequency. The plot was calculated by keeping the two damping rates fixed ($z_q$= 1 and $z_{\sigma}$= 1.05), while the ratio of $f_{un}/z_q$ was varied as indicated on the horizontal axis. Small values of this ratio correspond to the hydrodynamic region. The other three coupling parameters were varied alongsize with this ratio such that at any value of  $f_{un}/z_q$ the following holds: $f_{un}= f_{uT}= f_{Tq}/1.5$, whereas the ratio $f_{u\sigma}/f_{un}$ is shown in the figure. In all the figures we have taken $z_{q\sigma}$= 0. The propagation frequencies (when unequal to zero) are shown in the left column, the damping rates in the second column, the amplitudes of the Lorentzian lines in the third column, and a bird's eye view of $S_{sym}(q,\omega)$ is shown in the rightmost column. The stars indicate that the variables have been scaled to $z_q$, such as $\Gamma^*=\Gamma/z_q$.We can see a transition from hydrodynamic behavior at small $f/z$ values (two propagating modes and one central line) to more complicated spectra such as the presence of two pairs of propagating modes. The dotted line for the propagation frequencies represents the hydrodynamic sound velocity $[f_{un}^2+f_{uT}^2]^{0.5}$ while the hydrodynamic damping rate $f_{Tq}^2/z_q$ is given by the dashed-dotted curve in the second column. Note the rapid change in amplitude of the various lines and mixing of the modes while the underlying coupling parameters only vary gradually. More examples of this behavior are shown in Appendix D of Ref. \cite{wouterbook}.} \label{oup}
\end{figure*}

Hence, the recipe for doing a fail-safe fit is rewriting Eq. \ref{barocchi} as a sum of pairs given by Eq. \ref{anytwo}, add an isolated central Lorentzian to the summation for the case of an odd total number of Lorentzian lines (which will always result in at least one central line), and fit the data to sets of real parameters ($A,B,\Omega,z$). Doing so we obtain our (master) equation Eq. \ref{failsafe}. Thus, every single Lorentzian line yields two independent fitting parameters. As mentioned, the number of independent parameters can be reduced by imposing frequency sum rules.\\

The poles of the dynamic susceptibility can be determined through a fitting procedure to Eq. \ref{failsafe}. However, interpreting what these poles mean is not as straightforward. Both the numerical values of the poles, as well as their character can change rapidly as a function of $q$. For instance, when a mode goes from damped to overdamped, indicated by a rapid variation of the fitting parameters $A$ and $B$, then this might indicate a change through which decay channel density fluctuations relax. However, it does not imply that the coupling parameters or the underlying damping rates change rapidly. We illustrate this in Fig. \ref{oup} where we have slowly varying coupling parameters and unvarying damping rates that result in very rapid variations in the dynamic structure factor and in the poles of the dynamic susceptibility. In fact, the amplitudes of the Lorentzian lines vary so rapidly in certain q-ranges (Fig. \ref{oup}) that the parameters resulting from fitting to Eq. \ref{failsafe} will display discontinuous behavior.\\

Should we find ourselves in the situation where we run into the occurence of a rapid variation of our fit parameters, then we can try an equivalent, but more stable approach. In order to get a better handle on the character of the excitations and the physical processes behind them, we could try to populate an $m$ x $m$ matrix (representing $m$ Lorentzians) and redo the fit. Then we could remove parameters from the matrix until it affects the quality of the fit. Once we have obtained the matrix with a minimum number of parameters that still produces a fit comparable in quality to the one we obtained from Eq. \ref{failsafe}, then we can construct a decay tree and start to identify the character of the new type of excitation and exactly how it couples to the decay of longitudinal density fluctuations. This process should make clear how the new excitations couple to the standard decay trees of liquids, and of course, whether such couplings would be allowed on physical grounds. Next, we would compare this to computer simulations. The main advantage of populating a matrix is that, unlike the poles of a sum of Lorentzian lines, the parameters of the matrix only vary slowly with $q$. This is essentially the way the modes in normal liquids were identified, and the way in which the description of binary mixtures was put on a solid theoretical footing\cite{schep,westerhuis}. The most important point though in performing fits is to ensure that we fit to the type of expression given in Eq. \ref{failsafe} rather than putting in beforehand whether modes are propagating, or not, nor whether they are symmetric, or not.\\

In summary, in this paper we have focused on the details of the various models that are being employed to describe the scattering by liquids. We have discussed the formal identity between the projection formalism, the memory formalism, the continued fraction expansion, and the sum of Lorentzian lines when evaluated in the knowledge of the Liouville operator. We have stressed how these exact models can be (accidently) turned into phenomenological models through being too restrictive, or not restrictive enough. We have identified some common mistakes that can be found in the literature in the actual application of these models, such as generalizations that turn out to be inconsistent with theory.\\

We believe that the accuracy of recent scattering data demand that we do not employ (accidental) phenomenological models until we know for sure that exact formalism do not give an adequate fit to the data. Once we have reached that point then by using Eq. \ref{failsafe} we can be confident that when modeling of novel excitations we are not identfying features that are an artifact of the model being fitted to.  The latter case is not only relevant to transverse modes that have been observed in liquids, but also to the case where the electronic degrees of freedom couple to the ionic degrees of freedom on the time-scales probed by the experiment.\\


\begin{thebibliography}{10}
\bibitem{adhoc} bunch of papers that use delta function
\bibitem{never} Rather than listing all the scattering papers that ever have been published that contain results for $S(q)$ based on model fits, the reader can choose any of the experimental papers already listed under the references to get the gist of it.
\end{thebibliography}
\end{document}